%% file: main.tex
\documentclass[journal]{IEEEtran}
\ifCLASSINFOpdf
  \usepackage[pdftex]{graphicx}
\else
\fi
\usepackage{array}
\usepackage[caption=false,font=footnotesize]{subfig}
\usepackage{standalone}
\usepackage{float,adjustbox}
\usepackage{amsmath,amssymb}
\usepackage{amsthm}
\usepackage[margin=0.5in]{geometry}
\usepackage{color,soul}
\usepackage[usenames,dvipsnames]{xcolor}
\usepackage{enumerate}
\usepackage{epstopdf}
\usepackage{tikz}
\usetikzlibrary{shapes,arrows}
\usetikzlibrary{calc,fit,arrows}
\usetikzlibrary{decorations.pathmorphing} % noisy shapes
\usetikzlibrary{backgrounds}	% drawing the background after the foreground
\restylefloat{figure}

\newtheorem{prop}{Proposition}

\newtheorem{definition}{Definition}
\newtheorem{corol}{Corollary}
\newtheorem{lem}{Lemma}
\newcommand{\objfC}{log\left(\frac{1+S_x(f)|H(f)|^2}{1+S_x(f)|H(f)|^2e^{-C(f)}} \right)}
\newcommand{\bbf}[1]{\mathbf{#1}}

\newcommand{\intw}[1]{\int_{0}^{W}{#1}df}

\newcommand{\beqd}[1]{\begin{equation}#1\end{equation}}
\newcommand{\beqdn}[1]{\begin{equation}\nonumber#1\end{equation}}
\newcommand{\dd}[2]{\frac{\partial #1}{\partial #2}}
\newcommand{\Xf}{H(f)^2-\lambda _s -\lambda _c H(f)^2}

\newcommand{\nfigSing}[3]{
  \begin{figure}[t]
  \centering
\includegraphics[trim=0cm 0cm 0cm 0cm,clip=false,width=4in,height=3in]{#1}
 \caption{#2} 
 \label{#3}
 \end{figure}
}
\begin{document}
%
% paper title
% Titles are generally capitalized except for words such as a, an, and, as,
% at, but, by, for, in, nor, of, on, or, the, to and up, which are usually
% not capitalized unless they are the first or last word of the title.
% Linebreaks \\ can be used within to get better formatting as desired.
% Do not put math or special symbols in the title.
\title{Oblivious Fronthaul-Constrained Relay for a Gaussian Channel}
%
%
% author names and IEEE memberships
% note positions of commas and nonbreaking spaces ( ~ ) LaTeX will not break
% a structure at a ~ so this keeps an author's name from being broken across
% two lines.
% use \thanks{} to gain access to the first footnote area
% a separate \thanks must be used for each paragraph as LaTeX2e's \thanks
% was not built to handle multiple paragraphs
%

\author{Adi~Homri,~\IEEEmembership{Member,~IEEE,}
        Michael~Peleg,~\IEEEmembership{Member,~IEEE,}
        and~Shlomo~Shamai,~\IEEEmembership{~Fellow,~IEEE}% <-this % stops a space
\thanks{This research has received funding from the European
Union's Horizon 2020 Research And Innovation Programme under grant
agreement no. 694630.}% <-this % stops a space
}

% note the % following the last \IEEEmembership and also \thanks - 
% these prevent an unwanted space from occurring between the last author name
% and the end of the author line. i.e., if you had this:
% 
% \author{....lastname \thanks{...} \thanks{...} }
%                     ^------------^------------^----Do not want these spaces!
%
% a space would be appended to the last name and could cause every name on that
% line to be shifted left slightly. This is one of those "LaTeX things". For
% instance, "\textbf{A} \textbf{B}" will typeset as "A B" not "AB". To get
% "AB" then you have to do: "\textbf{A}\textbf{B}"
% \thanks is no different in this regard, so shield the last } of each \thanks
% that ends a line with a % and do not let a space in before the next \thanks.
% Spaces after \IEEEmembership other than the last one are OK (and needed) as
% you are supposed to have spaces between the names. For what it is worth,
% this is a minor point as most people would not even notice if the said evil
% space somehow managed to creep in.

% The paper headers
\markboth{Journal of \LaTeX\ Class Files,~Vol.~14, No.~8, August~2015}%
{Shell \MakeLowercase{\textit{et al.}}: Bare Demo of IEEEtran.cls for IEEE Journals}
% The only time the second header will appear is for the odd numbered pages
% after the title page when using the twoside option.
% 
% *** Note that you probably will NOT want to include the author's ***
% *** name in the headers of peer review papers.                   ***
% You can use \ifCLASSOPTIONpeerreview for conditional compilation here if
% you desire.

% If you want to put a publisher's ID mark on the page you can do it like
% this:
%\IEEEpubid{0000--0000/00\$00.00~\copyright~2015 IEEE}
% Remember, if you use this you must call \IEEEpubidadjcol in the second
% column for its text to clear the IEEEpubid mark.

% use for special paper notices
%\IEEEspecialpapernotice{(Invited Paper)}

% make the title area
\maketitle

% As a general rule, do not put math, special symbols or citations
% in the abstract or keywords.
\begin{abstract}
We consider systems in which the transmitter conveys messages to
the receiver through a capacity-limited relay station.
The channel between the transmitter and the relay-station is assumed to
be a frequency selective additive Gaussian noise channel. It is assumed that the transmitter
can shape the spectrum and adapt the coding technique so as to optimize
performance.
The relay operation is oblivious (nomadic transmitters),
that is, the specific codebooks used are unknown.
% , while the spectral shape of the transmitted signal
% is available.
We find the reliable information rate that can be
achieved with Gaussian signaling in this setting, and to that end, employ Gaussian bottleneck results %\cite{bib.gib_information} 
combined with Shannon's incremental frequency
approach. 
%\cite{bib.shannon_basic}
We also prove that, unlike classical water-pouring, the allocated spectrum (power and bit-rate) of the optimal solution could frequently be discontinuous. These results can be applied to a MIMO transmission scheme. We also investigate the case of an entropy limited relay. We present lower and upper bounds on the optimal performance (in terms of mutual information), and derive an analytical approximation. 
\end{abstract}

% Note that keywords are not normally used for peerreview papers.

% For peer review papers, you can put extra information on the cover
% page as needed:
% \ifCLASSOPTIONpeerreview
% \begin{center} \bfseries EDICS Category: 3-BBND \end{center}
% \fi
%
% For peerreview papers, this IEEEtran command inserts a page break and
% creates the second title. It will be ignored for other modes.
\IEEEpeerreviewmaketitle
\input{introduction}

\input{system_model}

\input{quantization_alternatives}
\input{discrete_case_introduction}

\input{infinite_processing_time}

\input{finite_entropy}
\input{details_and_proves}

\input{conclusions}
\input{appendix}
\bibliographystyle{IEEEtran}
% argument is your BibTeX string definitions and bibliography database(s)
%\bibliography{IEEEabrv,../bib/paper}
%
%\bibliographystyle{unsrt}
%\bibliography{./bibtex/bib/michaelbilb}
% Generated by IEEEtran.bst, version: 1.14 (2015/08/26)

% biography section
% 
% If you have an EPS/PDF photo (graphicx package needed) extra braces are
% needed around the contents of the optional argument to biography to prevent
% the LaTeX parser from getting confused when it sees the complicated
% \includegraphics command within an optional argument. (You could create
% your own custom macro containing the \includegraphics command to make things
% simpler here.)
%\begin{IEEEbiography}[{\includegraphics[width=1in,height=1.25in,clip,keepaspectratio]{mshell}}]{Michael Shell}
% or if you just want to reserve a space for a photo:

\begin{IEEEbiography}{Adi Homri}
Biography text here.
\end{IEEEbiography}

% % if you will not have a photo at all:
% \begin{IEEEbiographynophoto}{John Doe}
% Biography text here.
% \end{IEEEbiographynophoto}

% % insert where needed to balance the two columns on the last page with
% % biographies
% %\newpage

% \begin{IEEEbiographynophoto}{Jane Doe}
% Biography text here.
% \end{IEEEbiographynophoto}

% You can push biographies down or up by placing
% a \vfill before or after them. The appropriate
% use of \vfill depends on what kind of text is
% on the last page and whether or not the columns
% are being equalized.

%\vfill

% Can be used to pull up biographies so that the bottom of the last one
% is flush with the other column.
%\enlargethispage{-5in}

% that's all folks
\end{document}

%% file: introduction.tex
\section{Introduction}

\IEEEPARstart{R}{elaying} exploits intermediate nodes to achieve communication between two distant nodes. Elementary relaying can be coarsely divided into compress-and-forward (of which amplify-and-forward is viewed as a special case) and decode-and-forward, 
 depending on whether the relays decode the transmitted message
 or just forward the received signal to the destination.
In this paper we examine the ``oblivious'' relay system. The
oblivious approach constructs universal relaying components serving many diverse users and
operators and is not dependent on a priori knowledge of the modulation method and coding. This approach 
might benefit systems used in 'cloud' communication and was investigated, for example, in \cite{bib.dist_mimo_upper}. 

Consider the system in Fig. \ref{fig:system_model}. The information source $U,\;H(U)\;\leq R\text{[nats/sec]}$ is encoded into Gaussian symbols $X$ and transmitted via a Gaussian scalar channel; the relay compresses
the received symbols, $Y$, encodes them into a bit-stream $\bbf{B}$ and forwards it (without errors) to the final user's destination by a finite rate link $H(\bbf{B})\;\leq C\;\text{[nats/sec]}$. At the destination, a decompressor decodes the bit-stream into symbols $Z$ which are now an input for the receiver for estimation of $U$ i.e. $\hat{U}$. 

For the user, the relay operation is hidden as it transmits the symbol $X$ and receives
symbol $Z$, while the effective channel is governed by the transition probability $P_{Z|X}(z|x)$. This setting provides the user a memoryless communication channel that forwards
symbols from the transmitter to the receiver. 
We choose $X$ to be Gaussian because of its optimality subject to a large bit-rate constraint $C$ and because of its ubiquitous applications.
In this setting, the user faces the familiar memoryless communication channel
and can choose freely how to utilize it, e.g. the user can select a good error correcting
code and change the codes after the oblivious system was already implemented.
The serving system is oblivious of the channel code used (see
\cite{DcntrlizedProc} for a more rigorous presentation of obliviousness).
The relay performs lossy compression of the output of the Gaussian channel and is implemented by source coding. 
The trade-off between compression rate and mutual information between channel input
and compressed channel output has closed-form expressions for the scalar and
vector case using the Gaussian Information Bottleneck (GIB) theorem  \cite{bib.gib_information,bottleneck_explain} and
\cite{bib.bottelneck_gaussian_vector}.
This deviates from the classical remote rate-distortion approach 
\cite{bib.berger}, \cite{bib.Doburshin_Tsybakov}, \cite{bib.kipnis_et_al} (rate distortion for sub-Nyquist sampling scheme) and \cite{DBLP:journals/corr/KipnisEG16} (sampling stationary signals subject to bit-rate constraints), since the distortion is measured by the equivocation $h(X|Z)$ instead of by the $MMSE=E(X-Z)^2$. Since the distribution of $X$ is fixed, minimizing $h(X|Z)$ means maximizing $I(X;Z) = h(X) - h(X|Z)$. 

\definecolor{mycolor}{rgb}{0.6,0.6,1}
\tikzstyle{new_block} = [draw, fill=blue!10, rectangle,
minimum height=3em, minimum width=5em,text width = 5em]
\tikzstyle{chan_block} = [draw, fill=green!10, rectangle,
minimum height=3em, minimum width=4em,text width = 4em,text centered]
\tikzstyle{block} = [draw, text centered,fill=orange!30, rectangle,
minimum height=3em,minimum width=4em,text width = 4em]

\tikzstyle{mega_block} = [draw, fill=yellow!10, rectangle,
minimum height=7em, minimum width=25em]

\tikzstyle{pinstyle} = [pin edge={to-,thin,black}]
\begin{figure}[t]
\centering
\begin{tikzpicture}[auto, node distance=3cm,>=latex']
\node [coordinate, name=input] {};
\node [block,right of= input,xshift = 0cm] (enc) {Encoder};
% \node [block , right of=enc] (awgn) {Gaussian scalar channel}; 
\node[chan_block , right of=enc,name = awgn,xshift=0.5cm] {AWGN Channel};
\node [coordinate, right of=awgn,xshift = 0cm] (Ain) {};
\node [coordinate, below of=input,yshift = 0.5cm] (Aout) {};
\node [mega_block, right of=Aout,xshift = 1.75cm,yshift=0.2cm] (relayOp) {};
\node [new_block,below of = enc,yshift = 0.5cm](decomp){Decompress};
\node [new_block,below of = awgn,yshift = 0.5cm](comp){Compress};
\node [coordinate, below of=Ain,yshift = 0.33cm] (Bin) {};
\node [coordinate, below of=Aout,yshift = 0.35cm] (Bout) {};
\node [block,right of= Bout,xshift = 1.5cm,yshift = 0.5cm] (decoder) {Decoder};
 \node [coordinate, name=output,right of = decoder,xshift = 0.8cm] {};
% \node [new_block, right of=Bout,xshift = 0.2cm] (decomp) {Decompressor};
% 
%
\node at (relayOp.west) [above=9.5mm, right=0mm] {\textsc{Oblivious Constrained Relay}};

\draw [->] (input) -- node[midway,below] {} node[midway,above] {$U$}(enc);
\draw [->] (enc) -- node {$X$} (awgn);
\draw [-] (awgn) -- node {} (Ain);
\draw [->]  (Ain) |- node[near end,above] {$Y$} (comp.east);
\draw [->]  (comp.west) -- node [above]{$\bbf{B}$} (decomp.east) ;
\draw [-] (decomp.west) -- node [above] {Z} (Aout);
\draw [->]  (Aout) |- node {} (decoder.west) ;
% % \draw [-,dash dot] (Aout) |- node {$Y$} (decomp.west);
% % \draw [->,line width=1mm,  blue] (obv) -- node[midway,below] {}  node[midway,above] {$\bbf{B}$}(Bin);
% %\draw [->] (decomp) -- node {$Z$} (decoder); 
\draw [->] (decoder.east) -- node {$\hat{U}$} (output);
\end{tikzpicture}
\caption{The oblivious relay (blue) serving a user communicating via a Gaussian scalar channel (green).}
\label{fig:system_model}
\end{figure}
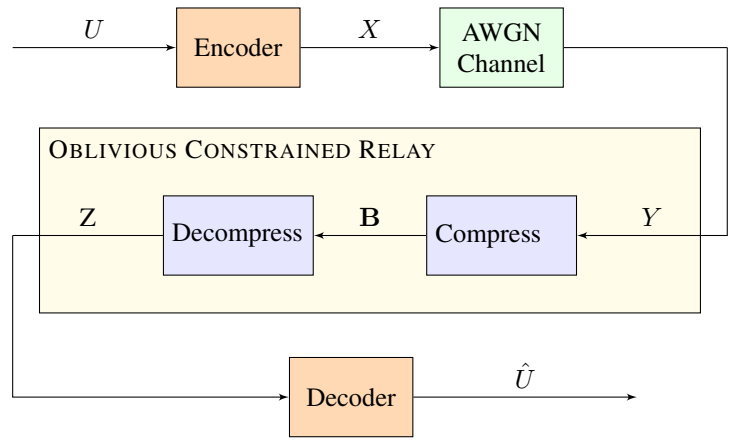

We further discuss the oblivious relay and focus our attention on the quantization process. We examine simpler quantizers which can be implemented by the standard Lempel-Ziv algorithm instead of source coding. The performance of such quantizers that are optimal relative to an entropy constraint was studied for a wide class of memoryless sources (e.g.  \cite{h.gishj.n.pierce1968},\cite{p.nollr.zelinski1978} and \cite{n.farvardinj.w.modestino1984}). Notwithstanding, it is interesting to investigate the effect of such a constraint on the relay operation. 

\textit{Our Contribution:} In this paper, we provide a further generalization of the GIB for the case of a
frequency selective additive Gaussian channel (some preliminary results were presented in~\cite{a.homrim.pelegs.shamai2016}). We find the reliable information rate that
can be achieved in this setting, and to that end employ Gaussian bottleneck results
\cite{bib.gib_information} combined with Shannon's incremental frequency
approach \cite{bib.shannon_basic}. The incremental approach leads to a clear solution for the frequency-selective channel setting. Analysis of frequency-flat channels and MMSE optimization was reported in \cite{siddhart2011} and \cite{siddhart2006}. Furthermore, in Section~\ref{sec:finite_entropy} we present lower and upper bounds for the mutual information between the transmitter and the receiver when the entropy constraint is placed on the relay.

The remainder of this paper is organized as follows: 
Section~\ref{sec:sys_model} provides the system model.
Section~\ref{sec:prelim} outlines preliminaries in which we summarize quantization
alternatives (\ref{sec:Quantization_alternatives}), demonstrate the advantages of stochastic quantizers (\ref{sec:advantage_stchostis_quantizer}), show that the optimal transmitting scheme dictates independent $z_i$ (\ref{sec:independent_zi_for_optimal_performance}),
provide the required background and definitions for the GIB (\ref{sec:GIB}) and review the classical water-pouring method (\ref{sec:classical_water_pouring}).
In Section~\ref{sec:inifinte_GIB}, we review the main results
relevant to frequency-flat channels from \cite{bottleneck_explain,bib.shannon_basic} and present the new derivation for frequency selective channels and
infinite-processing-time. Section~\ref{sec:finite_entropy} is dedicated to the finite entropy quantizer. Further derivations and proofs can be found in Section~\ref{sec:proves}. Conclusions and proposals for future work are found in Section~\ref{sec:conclusions}.

\textit{Notation:} $X$ is a random variable. $x$ is a realization of a random variable. We use boldface letters for column vectors and sequences. The
expectation operator is denoted by $E[\cdot]$ and we follow the notation of
\cite{bib.cover_and_thomas} for entropy $H(\cdot)$, differential entropy
$h(\cdot)$, and mutual information $I(\cdot;\cdot)$. A probability mass/distribution function is denoted by $P(\cdot)$ or $p(\cdot)$, respectively. All logarithms are natural and the unit of information is nats unless stated otherwise.

%% file: system_model.tex
\section{System Model}
\label{sec:sys_model}
\tikzstyle{block} = [draw, fill=orange!30,rectangle,
minimum height=4em, minimum width=6em]
\tikzstyle{gblock} = [draw, fill=green!30,rectangle,
minimum height=4em, minimum width=6em]
\tikzstyle{new_block} = [draw, fill=blue!10, rectangle,
minimum height=4em,text width=6em]
\tikzstyle{sum} = [draw, fill=white, circle, node distance=3cm]
\tikzstyle{input} = [coordinate]
\tikzstyle{output} = [coordinate]
\tikzstyle{pinstyle} = [pin edge={to-,thin,black}]
\newcommand{\suma}{$+$}
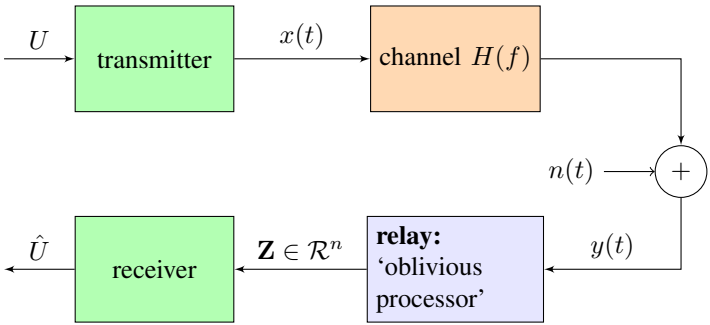
\begin{figure}[t]
\centering

\begin{tikzpicture}[auto, node distance=3cm,>=latex']
\node [input, name=input] {};
\node [gblock, right of=input,xshift=-1cm] (enc) {transmitter};
\node [block, right of=enc,xshift=1cm] (channel) {channel $H(f)$};
\node [sum, right of=channel,pin={[pinstyle,xshift=-0.2cm]left:$n(t)$},xshift=0cm,yshift=-1.5cm](sum){\suma}; 
%\node [below of=sum](switchpoint){};
\node [new_block, below of=channel,yshift=0.2cm,xshift=0cm] (obv) {\textbf{relay:} \qquad \qquad  `oblivious processor'};
\node [gblock, below of=enc,yshift=0.2cm] (dec) {receiver};
\node [output,left of=dec,xshift=1cm] (output) {};
\draw [->] (input) -- node {$U$} (enc);
\draw [->] (enc) -- node {$x(t)$} (channel);
\draw [->] (channel.east) -| node {} (sum);
\draw [->] (sum) |- node [near end,above]{$y(t)$} (obv);
%\draw [->] (switchpoint) -- node {}(obv);
\draw [->] (obv) -- node [above]{$\bbf{Z}\in\mathcal{R}^n$}(dec);
\draw [->] (dec) -- node [above]{$\hat U$} (output);
\end{tikzpicture}
\caption{A finite rate relaying operation over a fronthaul AWGN frequency selective channel.}
\label{system model waveform}
\end{figure}

Consider the system depicted in Fig. \ref{system model waveform}. $x(t)$ is the input signal, assumed to be Gaussian, $H(f)$ is the frequency response of the channel linear filter, and the impulse response $\mathcal{F}^{-1}\left[H(f)\right] = h(t)$ (here,
$\mathcal{F},\mathcal{F}^{-1}$ designate the Fourier response and its inverse)
\beqdn{y(t) = x(t)*h(t)+n(t),}
where $n(t)$ is normalized additive white Gaussian noise with one-sided power
spectral density $N_0 = 1[Watt/Hz]$, and $*$ designates convolution.
We are interested in the normalized mutual information when standard coding theorems \cite{bib.gallager} guarantee that the associated rate
can be reliably transmitted through the system
\beqd{\lim_{T\rightarrow
\infty}\frac{1}{2T}I\left(X_{-T}^{T};\bbf{Z}\right)\triangleq I_n^C(X;Z).
\label{eq:def_info_measure}} 
Denote $X_{-T}^T$ as $(X(t),-T\le t\le T)$, $\bbf{Z}$ is the output vector (containing the compressed channel outputs $Z_i$),  which is entropy constrained by $H(\bbf{Z})\leq nC$[nats/sec], The information in
(\ref{eq:def_info_measure}) is also measured in terms of [nats/sec]. $n$ denotes the number of symbols in a transmitted block and is of the dimension of $\bbf{Z}$.   
Again, we seek the (one-sided) power spectral density of the input
Gaussian process $S_x(f)$ which maximizes $I_n^C(X;Z)$ under an average power constraint in some bandwidth $W$:
\beqdn{\intw{S_x(f)} \le P.}

%% file: quantization_alternatives.tex
\section{Preliminaries}
\label{sec:prelim}
\subsection{Quantization Alternatives}
\label{sec:Quantization_alternatives}
Denote by $X, Y, Z$ the channel input, the received signal and the quantized
output, respectively. Our system will try to maximize $I(X;Z)$  which clearly
determines the maximal information rate $R$ of the whole system if the user utilizes good error correcting codes, while minimizing 
the bit-rate of the sequence $\mathbf{B}$. Here we list some possible approaches
to quantization:

\subsubsection{Using the channel code in the serving subsystem}
If the serving subsystem would not be oblivious, it would decode the original
information ($\hat{U}$) and send it as the sequence $\mathbf{B}$. In this case $R = min(C,
\text{radio channel capacity})$ would be achieved. But in this work the serving
subsystem is oblivious.

% \subsection {Scalar entropy constraint quantizer}
% The received signal $y$ is mapped to a discrete valued variable $z$ by a
% deterministic function. The function is optimized for Mutual Information (MUI),
% $I(x;z)$ per symbol with a constraint on the number of bits required to
% represent the quantizer output symbol.

\subsubsection{MUtual Information Constrained - Stochastic Quantizer (MUIC-SQ)} 
A class of oblivious quantizers is stochastic, as mentioned for example in \cite{bib.inform_bottel_tishby}. For each channel output $Y_i$, a compressed representation $Z_i$ is obtained by a stochastic quantizer characterized by the probability mass function $P_{Z|Y}(z_i|y_i)$ chosen to maximize $I(X_i;Z_i)$; then $Z_i$ is compressed and sent to the user's decoder using the bit rate $C= I(Y_i;Z_i)$.
The practical implementation is by means of source coding on sequences. The received sequence $\bbf{Y}$ is encoded into the sequence of bits $\bbf{B}$ 
and the destination recovers the sequence $\bbf{Z}$ from $\bbf{B}$. The serving system bit rate can be limited to $C= I(Y_i;Z_i)$. A proof following the steps of the source coding theorem \cite{bib.cover_and_thomas} can be constructed. The probability mass function $P_{Z|Y}(z_i|y_i)$ will set $I(X_i;Z_i)$ and, thus, enable a system communication rate of $R = I(X_i;Z_i)$ by the classic channel coding theorem.

Letting the quantizer be stochastic improves performance, similarly to a corresponding advantage of source coding over memoryless deterministic quantization \cite{bib.cover_and_thomas}. In \cite{bib.koch}, Koch treats a stochastic quantizer, where the randomness is limited to dither known to the quantizer; this is a special case and may be considered a deterministic time-varying quantizer. 

The optimal stochastic quantizer for Gaussian signals is the GIB, and was thoroughly analyzed in~\cite{bottleneck_explain} and~\cite{bib.optimal_gib_curve}.

The GIB is a corner stone in this paper and its attributes will be specified in the upcoming section.
 
\subsubsection{Entropy Constrained Stochastic Quantizer (EC-SQ)}  
The entropy constrained stochastic quantizer (EC-SQ), works in the same way as MUIC-SQ, except that the entropy of the compressed channel output $Z_i, H(Z_i)$ is bounded to be less than $C$. Entropy compression schemes such as Huffman or Lempel-Ziv are added after the quantizer, as suggested in the literature.

It is evident that in terms of mutual information $I(X_i;Z_i)$, the EC-SQ is inferior to MUIC-SQ since 
% \begin{align}
% \label{eq:gib_upper_bound} 
% \max_{P(z_i|y_i) \;:\; H(z_i)\le C} I(x_i;z_i) & \leq 
% \max_{P(z_i|y_i) \;:\; I(y_i;z_i)\le C} I(x_i;z_i)
% \end{align}
\beqd{
\label{eq:gib_upper_bound} 
I(Y_i;Z_i) = H(Z_i)-H(Z_i|Y_i) \leq H(Z_i),
}
thus enforcing a tight constraint on $P_{Z|Y}(z_i|y_i)$.
For the Gaussian case the upper bound is the $I_{GIB}$.

\subsubsection{Entropy Constrained Deterministic Quantizer (EC-DQ)}
This quantizer assumes deterministic mapping $Z_i = f(Y_i)$ (i.e. $H(Z_i|Y_i) = 0$), where $f(\cdot)$ is some function on the channel output $Y_i$. It is clear that for general channels it is inferior to the EC-SQ, as the deterministic domain is a subset of the stochastic domain. In the AWGN channel, there is a deterministic quantizer with identical performance. This can be proven using the following steps:
\begin{itemize}
\item Split the range of the channel output $Y_i$ into small segments.
\item Perform a hair splitting operation on each segment in order to have \textit{deterministic} mapping that would yield the desired transfer function from $Y_i$ to $Z_i$.
\end{itemize}
See rigorous proof in Appendix~\ref{sec:equivalnce_stochastic_deterministic_quantizers}.

% Assume we have the entropy constrained stochastic quantizer; that is, we have $P(z_i|y_i)$. We now perform a hair splitting operation in which for \textit{each} possible $z_i = \eta$ we attach set of $\{y _l\}$, 
% \beqdn{y_i-\epsilon\leq y _l \leq y_i+\epsilon ,\; \epsilon \geq 0}.
% and form a deterministic quantizer: $z_i = \eta$ if $y_i \in \{y _l\}$ with the same performance as the stochastic one. The set $\{y _l\}$ should be chosen with care in order to maintain $P_{Z_i|Y_i}(z_i = \eta |y_i) \approx P_{Y_i}(\{y _l\})$

\subsubsection{Memoryless deterministic quantizer}
The received signal $Y$ is mapped to a discrete valued variable $Z$ by a
deterministic function. The function is optimized for mutual information
$I(X;Z)$ per symbol with a constraint on the number of bits, or alphabet size of $Z$, required to
represent the quantizer output symbol. The optimization can be done by the Lloyd
algorithm. This is well covered by published papers,
e.g.~\cite{bib.cap_under_output_quant}, which also show that the optimal probability distribution function of the transmitted signal $X$ is  discrete in many cases.

\subsubsection{Vector quantizers} 
Assume a vector compression scheme in which we group a few variables $Z$ into small $n$ -length vectors $\bbf{Z_k} = \left[Z_{k+1} ,Z_{k+2}, \ldots , Z_{k+n} \right]$, each being a deterministic function of the vector $\bbf{Y_k} = \left[Y_{k+1} ,Y_{k+2}, \ldots , Y_{k+n} \right]$ under the constraint
\beqd{H(\bbf{Z_k})\leq nC.}
Entropy coding will still be possible, now over $\bbf{Z_k}$
instead of the scalars $Z$. This possibility leads to
the following observations:
\begin{itemize}
\item With large $n$ we can implement the full GIB by compressing the sequence $\bbf{Y}$ into sequence $\bbf{Z}$ by the MMSE criterion under the constraint of bit-rate $C$. 
%\item The quantization schemes per symbol (stochastic entropy constraint, deterministic entropy constraint and GIB) will all lead to the same results when $n$ is large. 
\item Vector quantizers provide many
intermediate performance levels starting at the deterministic entropy-constrained quantizer $(n=1)$ and up to the
GIB quantizer.
\end{itemize}

% When vector compression is applied, the quantization schemes (stochastic quantizers entropy/mutual-information constraint, deterministic entropy constraint) will lead to the same results; rendering the GIB upper bound to be tight.

The advantage of the stochastic quantizer over the entropy constrained quantizer is the advantage of source coding over a scalar quantizer. Next, we shall present some attributes of the stochastic quantizer.

% which can be shown by the source coding theorem in \cite{bib.cover_and_thomas}
% %using a distortion measure defined as
% 
% \begin{equation} \label{eq:dist}
% d(y,z) = log\left(\frac{P(z_i|y_i)}{P(z_i)}\right)
% \end{equation}
% The rate distortion function D(R) of this special distortion measure is D=R.
% This follows from the properties of typical sequences, e.g.
% \cite{bib.cover_and_thomas} since:
% 
% \begin{equation}
% \frac{1}{N}\sum_{i=1}^N log\left(\frac{P(z_i|y_i)}{P(z_i)}\right)\cong
% E\left[log\left(\frac{P(z_i|y_i)}{P(z_i)}\right)\right] = I(z_i,y_i)
% \end{equation}
% Thus the $C= I(yi;zi)$  rate of $\mathbf{b}$ is achievable. The distortion in \cite{bib.cover_and_thomas} is
% positive while in (\ref{eq:dist}) is not for all $z, y$, this can be corrected
% by adding a positive constant offset. A perfectly rigorous proof following the steps of
% the source coding theorem \cite{bib.cover_and_thomas} can be constructed.
% \hl{Adi, if you wrote one we can use it} The PDF $P(z_i|y_i)$  will set
% $I(x_i;z_i)$ and thus enable system communication rate of $R=I(x_i;z_i)$ 
% by the classic channel coding theorem.
\subsection{Demonstrating the advantages of the stochastic quantizer}
\label{sec:advantage_stchostis_quantizer}
\begin{figure}[!t]
\centering
\includegraphics[width=3.5in,height=3in]{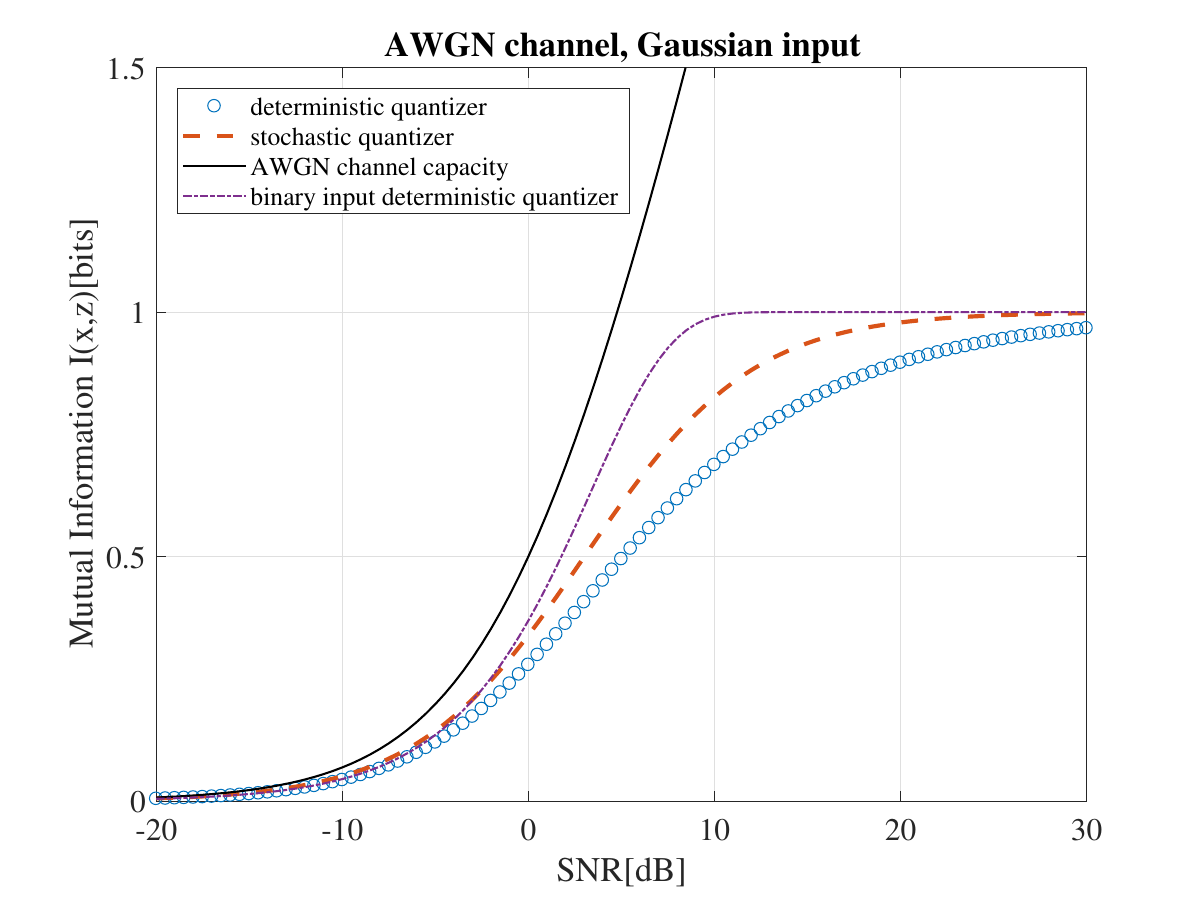}
\caption{Mutual information and system information rate $R$ with Gaussian signal
and two quantizers over an AWGN channel and with a quantization rate of
$C = 1$[bit/channel use] as a function of SNR. Binary input is also presented for comparison.}
\label{fig:performance_comprasion_awgn}
\end{figure}

The advantage of the stochastic quantizer is demonstrated by a numerical example, see Fig. \ref{fig:performance_comprasion_awgn}.  We examine the case of a Gaussian $X$ over an AWGN channel with a quantization rate $C = 1$[bits/symbol].
In the memoryless deterministic quantizer case, the quantizer is the sign of the received signal. Using Kindler~\cite{g.kindlerr.o'donnelld.witmer2016}, the sign of the received signal is the optimal one bit memoryless deterministic quantizer, and not necessarily the optimal entropy constrained deterministic quantizer. The curve in
Fig.~\ref{fig:performance_comprasion_awgn} is the numerical evaluation of
$E_{X,Z}log\frac{P(z|x)}{P(z)}$. In the stochastic case we have, from \cite{bib.optimal_gib_curve}, the GIB
\beqdn{I(X;Z) = \frac{1}{2}log\left(\frac{1+SNR}{1+SNR\cdot e^{-2C}}\right). }
The results in Fig. \ref{fig:performance_comprasion_awgn} show the clear superiority of the stochastic quantization over the deterministic one.
% , as well as the optimality of a 1-bit transmission in a Gaussian channel at low snr, as depicted in \cite{g.kindlerr.o'donnelld.witmer2016}. 
Modifying the distribution of $X$ would improve the rate \cite{bib.cap_under_output_quant} (see the improved performance with
 a binary input in Fig. \ref{fig:performance_comprasion_awgn}).

\subsection{Independent $Z_i$ achieve the optimal performance}
\label{sec:independent_zi_for_optimal_performance}
We might still wonder if the stochastic quantizer,
while evidently better than the deterministic one, is optimal.
That is, the $Z_i$ in our scheme are statistically independent. Could this
scheme be outperformed if dependence between the $Z_i$ was permitted?
For example, the channel from $X$ to $Y$ could be a BSC and the relay could
convey information to the destination by setting $Z_i$ to be parity bits
obtained by the XOR operation on pairs of $Y_i$. We shall show next that the
independent $Z_i$, each statistically dependent on a single $Y_i$ only, achieve
the best performance possible. To show this, we consider the scheme as in
Fig.~\ref{fig:system_model}, but instead of producing $\bbf{Z}$, the
bit sequence $\bbf{B} = \bbf{Z}$ is derived directly from the sequence $\bbf{Y}$ and passed to the decoder
together with the compression scheme. Thus, we want to maximize $I(\bbf{X};\bbf{Z})$,
that is, the mutual information of whole sequences, with a constraint on $I(\bbf{Y};\bbf{Z})$. The
first term is the information rate of the whole system and the second term is an
achievable lower bound on the backhaul bit-rate C.
We can restate this question as an equivalent bottleneck problem:  Let
$\bbf{X,Y,Z}$ be sequences, each comprising $n$ elements $X_i, Y_i$ and $Z_i$. Also, let the
elements of $\bbf{X, Y}$ be i.i.d. and the channel $\bbf{X\;-\;Y}$  be
memoryless. In this case, the bottleneck problem is finding $P_{Z|Y}(\bbf{z|y})$ which maximizes $I(\bbf{X;Z})$ with a constrained
$I(\bbf{X;Y})$, and the question on hand is: 
\beqdn{
\emph{Is $P_{Z|Y}(\bbf{z|y}) = \Pi_iP_{Z|Y}(z_i|y_i)$ ?}
}
The answer was already proved positive by Witsenhausen and Wyner \cite{bib.hans} for discrete alphabets of $X, Y$ and also for a Gaussian $X$ over the AWGN channel in Tishbi \cite{bib.optimal_gib_curve} (An alternative proof for continuous alphabets is available in~\cite{DBLP:journals/corr/HomriPS15}).

%% file: discrete_case_introduction.tex
\subsection{Gaussian Information Bottleneck (GIB)}
\label{sec:GIB}
\subsubsection{Information rate - scalar channel}

The GIB and its derivation for the discrete-time signaling case was thoroughly studied in
\cite{bib.gib_information,bottleneck_explain,bib.bottelneck_gaussian_vector,bib.optimal_gib_curve}
and \cite{bib.ib_method}.
We will now give a brief overview of the GIB. The interested reader is referred
to \cite{bottleneck_explain,bib.bottelneck_gaussian_vector} for a full treatment. A complete derivation of the information
rate function for the vector case, as well as the difference between the information rate function and the
rate-distortion function, namely, $I(R)\ge I^{RD}(R)$, is presented in
\cite{bib.bottelneck_gaussian_vector}.

\tikzstyle{new_block} = [draw, fill=blue!10, rectangle,
minimum height=3em, minimum width=4em]
\tikzstyle{block} = [draw, text centered,fill=orange!30, rectangle,
minimum height=3em,minimum width=4em,text width = 4em]
\tikzstyle{sum} = [draw, fill=white, circle, node distance=3cm]
\tikzstyle{input} = [coordinate]
\tikzstyle{output} = [coordinate]
\tikzstyle{pinstyle} = [pin edge={to-,thin,black}]
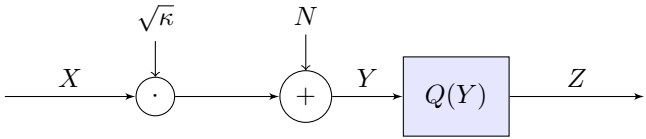
\begin{figure}[t]
\centering
\begin{tikzpicture}[auto, node distance=4.5cm,>=latex']
\node [input, name = pre_input]{};
\node [input, name=input,right of = pre_input,xshift = 2cm] {};
%\node [block,right of= input,xshift = -2.5cm] (chan) {\LARGE $\sqrt \kappa$};
% \node [circ, right of=enc,pin={[pinstyle]above:$\sqrt{a}$}] (channel)
% {$\times$}; 
% \node [circ, right of=channel,pin={[pinstyle]above:$n$}]
% (sum){$+$}; 
\node [sum, right of=input,pin={[pinstyle]above:$\sqrt\kappa$},xshift = -1cm]
(chan){$\bbf{\cdot}$}; 
\node [sum, right of=chan,pin={[pinstyle]above:$N$},xshift = -1cm]
(sum){$+$}; 
\node [new_block, right of=sum,xshift = -2.5cm] (obv)
{$Q(Y)$};
%{Oblivious
% Relay};
\node [output, name=output,right of = obv,xshift = -2cm] {};
\draw [->] (input) -- node {$X$} (chan);
\draw [->] (chan) -- node {} (sum);
% % % %\draw [->] (channel) -- node {} (sum);
\draw [->] (sum) -- node {$Y$} (obv);
\draw [->] (obv) -- node {$Z$} (output);
% \draw [->] (decomp) -- node {$\bbf{z}$} (decoder);
% \draw [->] (decoder) -- node {$\hat{u}$} (output);
\end{tikzpicture}
\caption{Gaussian Information Bottleneck.}
\label{fig:system_model2}
\end{figure}

\nfigSing{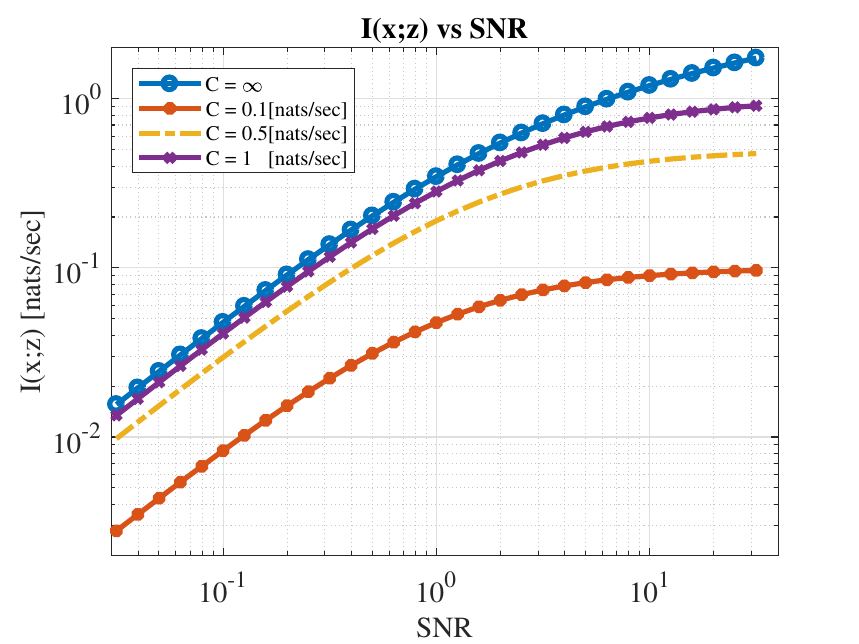}{GIB: Mutual information vs. rate($C$) and $\rho$(SNR).}{fig IvsR}
% \begin{figure}[!t]
% \centering
% \caption{$I(x;z)$ vs. Rate and $\rho$(SNR)}
% \label{fig IvsR}   
% \includegraphics[width=4in]{graphics/11}     
% \end{figure}

Consider the system in Fig. \ref{fig:system_model2}.  
% Let $\bbf{x}$ and
% $\bbf{y}$ be jointly Gaussian zero-mean random vectors(length $n$) with full rank covariance matrices.
% Assuming that $\bbf{x}\sim\mathcal{N}(\bbf{0,R_\bbf{xx}})$. The channel output
% equals
% \beqdn{\bbf{y=Hx+n}}
% where $\bbf{H}\in\mathcal{R}^{n \times m}$ and the additive noise
% $\bbf{n}\sim\mathcal{N}(\bbf{0},\sigma^2\bbf{I})$ is independent of $\bbf{x}$
% which results
% \beqdn{\bbf{R_{yy}} = \bbf{HR_{xx}H^T}+\sigma^2\bbf{I}}
% Let $\bbf{z}$ be a compressed representation of
% $\bbf{y}$ denoted by the conditional distribution $P(\bbf{z|y})$. It follows
% that $\bbf{x - y - z}$ forms a Markov chain and hence by Markovity
% \beqdn{P(\bbf{z|x}) = \int_{\mathcal{R}^n}P(\bbf{z|y})P(\bbf{y|x})d\bbf{y}}
% The compression rate equals $I(\bbf{z;y})$.
The GIB addresses the following variational problem \cite{bib.inform_bottel_tishby}: 
\beqdn{\min_{P(z|y)} I(Y;Z)-\beta I(X;Z).} 
In the context of the information bottleneck method, $X$ is called the 
\textit{relevance variable} and $I(X;Z)$ is termed \textit{relevant
information}. The trade-off between compression rate and relevant information is
determined by the positive parameter $\beta$. It has been shown that the
optimal $z$ is jointly Gaussian with $y$ and can be written as
\beqdn{Z = \alpha Y+\xi,}
where $\alpha\in\mathcal{R}$ is scalar and $\xi \sim
\mathcal{N}(0,\sigma_\xi)$ is independent of $Y$ . 
\begin{definition}
Let $X - Y -Z$ be a Markov chain. The information rate function
$I:\mathcal{R}_+ \rightarrow [0,I(X;Y)]$  is defined by \cite{bottleneck_explain}
\beqdn{I(C) \triangleq \max_{P(z|y)} I(X;Z) \text{ subject to }
I(Y;Z)\le C.}
\end{definition}
$I(C)$ quantifies the maximum amount of the relevant information that can be preserved when
the compression rate is at most $C$.

Let us present $I(C)$ for the channel depicted in Fig.~\ref{fig:system_model2}.
Since $X$ and $Y$ are real zero-mean jointly Gaussian random variables, they obey
\beqdn{Y = \sqrt{\kappa} X+N,}
where $h\kappa\in\mathcal{R}_+$ and $N\sim \mathcal{N}(0,\sigma^2)$ 
is independent of $X$. Setting $X\sim \mathcal{N}(0,P)$ yields
$Y\sim \mathcal{N}(0,\kappa P+\sigma^2)$. The compressed representation of $Y$ is
denoted $Z = Q(Y)$. By the Markovity of $X - Y - Z$ we have
\beqdn{P_{Z|X}(z|x) = \int_{\mathcal{R}}P_{Z|Y}(z|y)p_{Y|X}(y|x)dy,}
where $p_{Y|X}(y|x)$ is the transition probability distribution function of the Gaussian channel and $P_{Z|Y}(z|y)$
describes the compression mapping $Q$.
The capacity of the Gaussian channel $p_{Y|X}(y|x)$ with average power constraint $P$
and no channel compression equals \cite{bib.cover_and_thomas} (units are [nats/channel use]):
\beqdn{\mathcal{C}(\rho)\triangleq \frac{1}{2}log(1+\rho)}
with $\rho$ denoting the signal-to-noise ratio (SNR)
\beqdn{\rho \triangleq \frac{\kappa P}{\sigma^2}.}
The following corollary states a closed-form expression for the information rate
function and its properties  \cite[Theorem 2]{bottleneck_explain}.
\begin{corol}
The information rate function of a Gaussian channel with SNR $\rho$ is given by
\beqd{\label{eq:gib_cap}
I(C) = \frac{1}{2}log\left(\frac{1+\rho}{1+\rho
e^{-2C}}\right).}
$I(C)$ has the following properties:  
   \begin{enumerate}[(a)]
     \item $I(C)$ is strictly concave 
     \item $I(C)$ is strictly increasing in C
     \item $I(C)\le min\{C,\mathcal{C}(\rho)\}$
     \item $I(0) = 0$ and $\lim_{C\rightarrow\infty}I(C)= \mathcal{C}(\rho)$
     \item $\dd{I(C)}{C} = (1+e^{2C}\rho^{-1})^{-1}\le \dd{I(C)}{C}|_{C=0} =
     (1+\rho^{-1})^{-1}.$
   \end{enumerate}    
\end{corol}
The proof is found in \cite{bottleneck_explain}. It should be noted that it can
also be proved using the I-MMSE relation \cite[Chapter 5, Section
7.1.3]{bib.immse}.
Fig. \ref{fig IvsR} illustrates the effect of limited-rate processing. It is
clear that the total mutual information is upper bounded by the capacity for AWGN
channels derived by Shannon \cite{bib.shannon_basic}.

\subsection{Water-pouring}
\label{sec:classical_water_pouring}
We recall the classical water-pouring approach which yields the maximum
$I_n^\infty (x;z)$ for $C\rightarrow\infty$. The idea of splitting the channel into incremental bands appears in
\cite{bib.shannon_basic} and \cite{bib.cover_and_thomas}, where each incremental band of
bandwidth $df$ is treated as an ideal (independent due to Gaussianity) band-limited
channel with response $H(f)df$, and the result yields
\beqdn{
\lim_{C\rightarrow\infty} I_n^C(x;z) \triangleq I_n^\infty(x;z) =
\intw{log\left[1+S_x(f)|H(f)|^2\right]}.}
Optimizing this over $S_x(f)$ under the power constraint yields (using the
standard Euler-Lagrange method \cite{bib.gelfand})
\beqdn{\frac{|H(f)|^2}{1+S_x(f)|H(f)|^2} = \frac{1}{b}.}
Thus, the result is (see \cite[chapter 8]{bib.shannon_basic})
\beqdn{I_{water-pouring}\triangleq I_n^\infty(x;z) = \int_{\mathcal{B}}
log\left[b|H(f)|^2\right]df}
and the frequency region $\mathcal{B}$ is given by
\beqdn{\mathcal{B} = \{f:b-\frac{1}{|H(f)|^2}\ge 0 \}.}

%% file: infinite_processing_time.tex
\section{water-pouring with the optimal quantizer}
\label{sec:inifinte_GIB}
\subsection{Processing under limited bit-rate $C$}
As before, we adopt Shannon's incremental view, taking advantage
of the fact that disjoint frequency bands are independent under the Gaussian law and stationarity. Let $\frac{1}{2}C(f)$ designate the number of [nats/channel use] assigned for delivering (processing) the band $(f,f + df)$. Since we have $2\cdot df$
independent channel uses (Nyquist) per second, the total rate per second in each band is 
\beqdn{\frac{1}{2}C(f)2df = C(f)df}
and, hence,
\beqdn{\intw{C(f)} = C.}
Culminating this view and incorporating 
(\ref{eq:gib_cap}), we reach the equation (for simplicity we denote $S_x(f)$ as $S(f)$)
\begin{multline}
I[f,S(f),C(f)] = \nonumber \\ \intw{
\log \left[
\frac
{1+S(f)|H(f)|^2}
{1+S(f)|H(f)|^2e^{-C(f)}}
\right]
}, \nonumber 
\end{multline}
leading to the following optimization problem:
\begin{multline}
\label{eq:cont_prob}
\max_{S(f),C(f)} \intw{I[f,S(f),C(f)]} \\ 
s.t. \quad \intw{S(f)}=P ,\intw{C(f)} = C.
\end{multline}
The solution of Eq.~(\ref{eq:cont_prob}) follows the standard Euler-Lagrange \cite{bib.gelfand} reasoning.
To that end, we follow the notation presented in \cite{bib.gelfand}. $I[f,\hat{S},\hat{C}]
\triangleq \objfC$ is the mutual information spectral
density [nats/sec/Hz]. Also, $\hat{S} \triangleq S(f)$, $\hat{C}
\triangleq C(f)$. The Lagrangian is
\beqd{ 
\label{eq:lagrange} 
L\left[f,\hat{S},\hat{C}\right] = I[f,\hat{S},\hat{C}] - \lambda _c \cdot \hat{C} -
\lambda _s\cdot \hat{S},
}  
where $\{\lambda _c,\lambda _s\}\in\Re$ are Lagrange coefficient multipliers. Differentiating Eq.~(\ref{eq:lagrange}) with respect to $\hat{C},\hat{S}$
and letting 
\beqdn{\hat{Q} \triangleq \exp(-\hat{C}),}
\beqdn{X_f \triangleq \Xf,}
will lead to the following equation (see complete derivation in \ref{sec:deriv_of_two_sltns}):
\beqd{0=-H(f)^2(1-\lambda_c )\hat{Q}^2 + X_f\hat{Q} -\frac{\lambda_s\lambda_c}{1-\lambda_c }. \label{eq:quadQ}}
The quadratic equation~(\ref{eq:quadQ}) produces two curve sets $\{S_i(f),Q_i(f)\} , i\in\{1,2\}$, where $i$ is defined in Sec.~\ref{apndx:proof_of_concave_sign}.
\begin{prop}
\label{prop:take_concave_sol}
We can discard the $\{S_{2}(f),Q_2(f)\}$ solution, since for \textit{each} frequency, regardless of $H(f)$, $I[f,\hat{S},\hat{Q}]$ is not concave in the pair $\{S_{2}(f),Q_2(f)\}$.
% , meaning that one could improve the "optimal" $I[f,\hat{S},\hat{Q}]$ simply by splitting the incremental frequency band.
\end{prop}

A rigorous proof can be found within Sec.~\ref{apndx:proof_of_concave_sign}, where we derived that for each frequency $f$, the optimal values for $S(f),Q(f)$ are 
\begin{subequations}
\label{eq:solLFinal}
\begin{align}
   S(f) & =  
   \begin{cases}
   \frac{X_f+\sqrt{X_f^2-4H(f)^2\lambda _c \lambda _s}}{2H(f)^2\lambda _s}    & f\in\mathcal{B}_l \\
   0 & f\not\in\mathcal{B}_l
   \end{cases},   
   \\    
   Q(f) & =  
   \begin{cases}
   \frac{X_f-\sqrt{X_f^2-4H(f)^2\lambda _c \lambda _s}}{2H(f)^2(1-\lambda _c)}    &  f\in\mathcal{B}_l\\
   1 & f\not\in\mathcal{B}_l
   \end{cases},    
\end{align}
\end{subequations}
where $\mathcal{B}_l$ is the set of frequencies that have non-zero resource allocation (bit-rate and power). In general, $\mathcal{B}_l$ is unique unless the channel has a flat sub-band response. The algorithm for constructing $\mathcal{B}_l$ can be found in Sec.~\ref{alg:setting_bl}.

In order to find the appropriate values for $\{\lambda _c,\lambda _s\}$ we had to use a grid search and the following proposition was used:
\begin{prop}
\label{prop:bound_lambda}
$\{\lambda _c,\lambda _s\}$ are bounded by
\beqd{0\le \lambda _s\le \max{H(f)^2}\quad , \quad 0\le \lambda _c\le 1 .}
\end{prop}
\begin{IEEEproof}
See Sec.~\ref{subsec:bound_lemma} for a rigorous proof.
\end{IEEEproof}

In stark contrast to classical water-pouring~\cite{DBLP:journals/corr/KipnisEG16} and \cite{bib.cover_and_thomas}, the optimal solution will frequently be discontinuous. As shown in Fig.~\ref{fig:unified_plot_concave_psi}, zero resources is a singular point inside the non-concave region. Since $C(f)$ and $S(f)$ can never drop gradually down to zero, the transition will always have an abrupt part. 

A simple example is the case where H(f) is constant over $f$, the SNR is sufficient and C is rather low. In this case an attempt to use frequency-constant $S(f)$ and $C(f)$ will place us in the non-concave region; a better solution will use only part of the available spectrum and utilize the available nats better by transmitting less information about the channel noise (see similar behavior in~\cite{siddhart2011}). 
% \begin{figure}[t]
% \centering
% \includegraphics[trim=0cm 0cm 0cm
% 0cm,clip=false,width=3.8in]{graphics/uniform_chan}
% \caption{Information rate as a function of allocated bandwidth}
% \label{fig:MUIvsB}
% \end{figure}
Fig. \ref{fig:MUIvsB} demonstrates this idea, assuming a flat channel (i.e. $H(f) = 1$). For a given total power $P = 2[Watt]$ , capacity $C = 0.5[nats/sec]$ and allocated user's bandwidth $W = 100[Hz]$, we calculated the mutual information rate when distributing the power and bit-rate uniformly over the bandwidth $B$ used:
\nfigSing{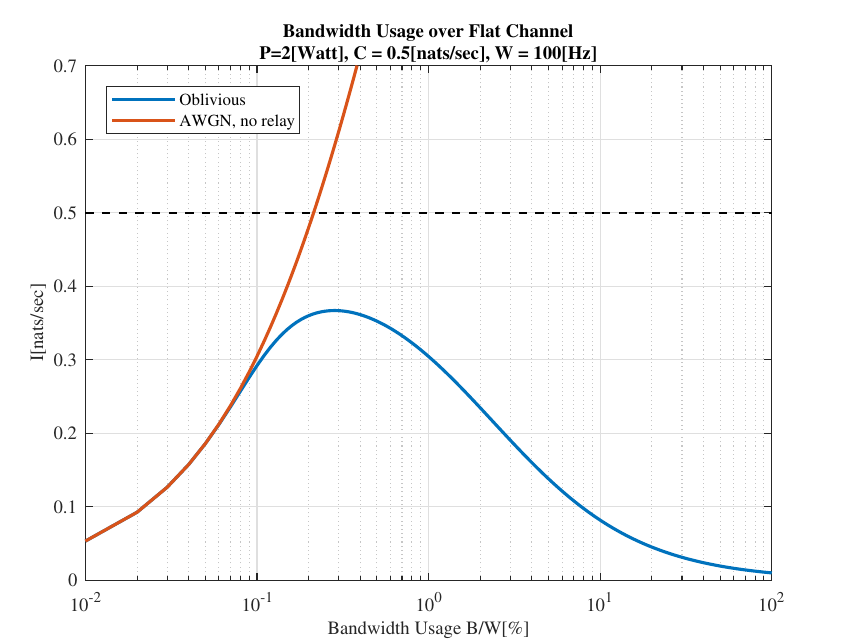}{Information rate as a function of allocated bandwidth.}{fig:MUIvsB}
\beqdn{
I \left[ S(f),C(f) \right] = B\log\left[\frac{1+P/B}{1+P/Be^{\frac{-C}{B}}}\right]
.}
It is clear that the best course would be to use only part of the spectrum, namely $B/W \approx 0.3[\%]$, which is the maximum of the oblivious curve (blue). 
\subsection{Numerical Analysis}
\label{sec:simulation}
%\subsection{Infinite processing time}

% \nfig{graphics/3}{graphics/4}{Allocated power and bit-rate vs arbitrary channel (a), and comparison to the allocated power resulted from water-pouring method (b), user bandwidth $W = 10[Hz]$}{multi_inverse}{!t}
% \nfig{graphics/9}{graphics/10}{An example of the abrupt nature of the optimal spectral allocation of power and bit-rate vs flat channel (a), and comparison to the allocated power resulted from water-pouring method (b), user bandwidth $W = 10[Hz]$}{flat}{!t}
\begin{table*}[!t]
\begin{center}
    \begin{tabular}{ | c | c | c | c | c | c | c |  p{2cm}| p{2cm} |c |}
    \hline
    Case & $\alpha _1 $ & $f_1$ & $\alpha _2 $ & $f_2$ & Remark & $I_n^C(x,z)$ & $I_n^C(x,z)$ - Uniform
    Allocation &  $I_n^C(x,z)$ - Limited-Rate Water-Pouring & $I_n^\infty(x,z)$ 
    \\
    \hline 
    1 & 0.25 & 0.25$W$ & 0.75 & 0.75$W$ & 
    Channel A %, Fig. \ref{fig:multi}
    & 2.94 & 2.83 & 0.77 & 2.944\\ \hline
    2 & 0.25 & 0.25$W$ & 0.75 & 0.75$W$ & Channel B 
    , Fig. \ref{fig:multi_inverse} 
    & 3.40  & 1.98 & 3.03 & 4.53\\ \hline
    
    3 & 0 & 0 & 1 & 0.5W & Channel A 
    %, Fig. \ref{fig:uni}
    & 3.73 & 0.98 & 3.68& 3.98 \\ \hline
    4 & 0 & 0 & 1 & 0.5W & Channel B 
    %, Fig. \ref{fig:uni_inverse}
    & 4.98 & 3.28 & 4.62 & 7.85 \\ \hline
    5 & - & - & - & - & Allpass , Fig. \ref{fig:flat}& 7.92 & 7.75 & 7.75& 23.98 \\ \hline
    6 & - & - & - & - & Allpass, $W=100[Hz]$ & 7.92  & 4.39 & 4.39& 69.31\\ \hline
    \end{tabular}    
\end{center}
\caption{ Comparison Between Channels (all units are [nats/sec])}
\label{tbl:summary}
\end{table*}

The proposed method has been applied on different types of
channels (denoted as ``Channel A'') of the form $H_A(f) \equiv
\alpha_1N(f_1,1) + \alpha_2N(f_2,1)$. $N(\mu,\sigma^2)$ is the Gaussian
curve with $P = 100[Watt]$ and $C = 9[nats/sec]$, while $W = 10[Hz]$ is the allocated user bandwidth. We also tested the ``reciprocal'' channel - denoted as ``Channel B''
(i.e, $H_B(f) = max[H_A(f)] - H_A(f)$).
In each scenario, we compared the overall information rate using the
following methods:
%\begin{itemize}[noitemsep,nolistsep]
\begin{itemize}
\item The proposed method
\item Uniform allocation of rate and power
\item Classical water-pouring, as presented in \cite{bib.shannon_basic}, for the
case of $C\rightarrow\infty$
\item ``Limited-Rate Water-Pouring'', which is:
\begin{enumerate}
  \item Calculate $S(f)$ using the classical water-pouring approach.
  \item The allocated rate is: $ C(f) = \frac{C}{P}S(f)$.
\end{enumerate}
\end{itemize}
\bigskip
The results are summarized in Table~\ref{tbl:summary}.
Figs.~\ref{fig:multi_inverse} and ~\ref{fig:flat} contain curves of $S(f), C(f), H(f)$ normalized to a unity average, and also a (normalized) classical water-pouring power allocation curve, $S(f)_{Water-Pouring}$, for comparison to the proposed approach.

It should be noted that the curves $S(f),C(f)$ of Fig.~\ref{fig:flat} are not unique (algorithm dependent) since the channel has a flat response; however, the total mutual information is maximized nonetheless.  
\nfigSing{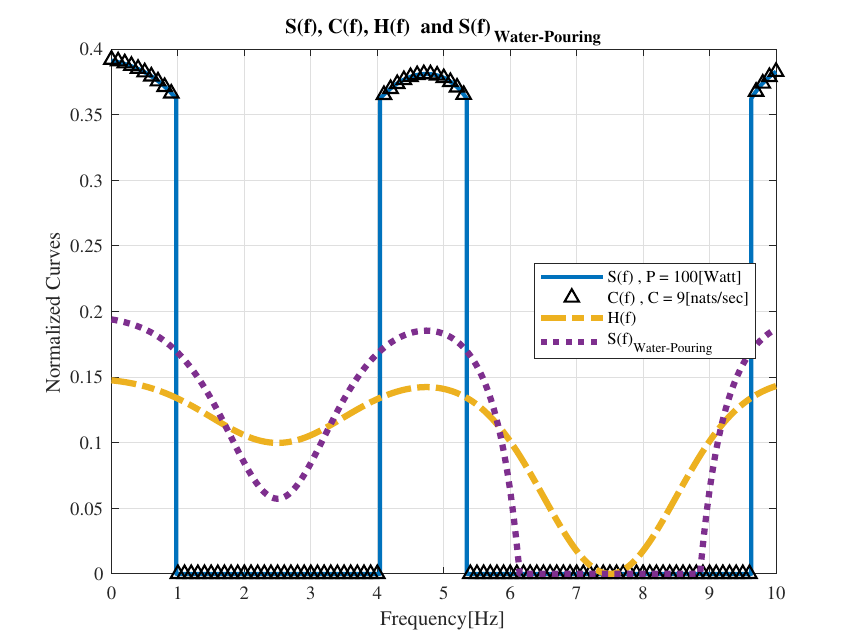}{Allocated (normalized) power and bit-rate vs arbitrary channel and comparison to the allocated power resulting from water-pouring method; user bandwidth $W = 10[Hz]$.}{fig:multi_inverse}
\nfigSing{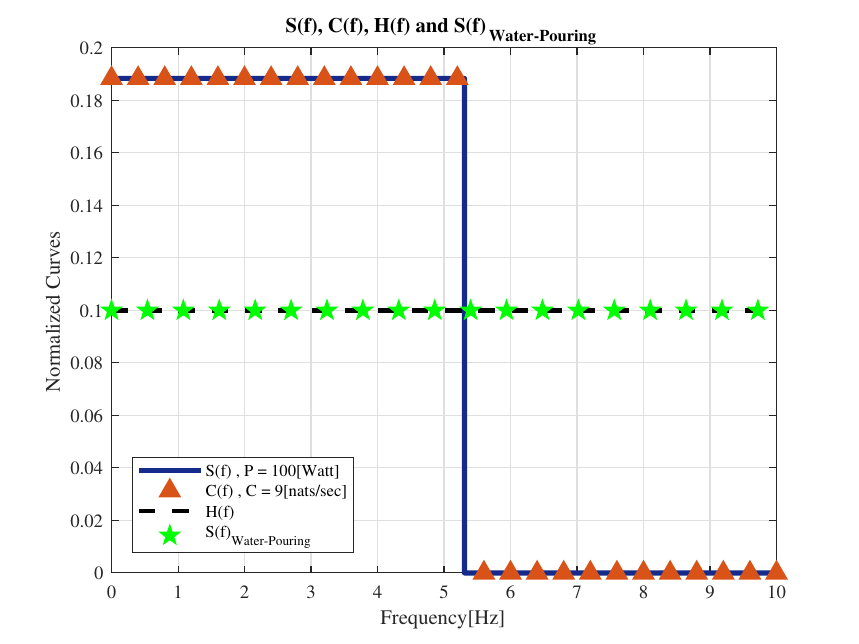}{An 
example of the abrupt nature of the optimal spectral allocation of power and bit-rate vs flat channel and comparison to the allocated power resulting from water-pouring method; user bandwidth $W = 10[Hz]$.}{fig:flat}
% \nfig{graphics/5}{graphics/6}{$\alpha_1 = 0,f_1 = 0, \alpha_2 = 1 , f_2 = 0.5W$
% }{uni}
% \nfig{graphics/7}{graphics/8}{$\alpha_1 = 0,f_1 = 0, \alpha_2 = 1 , f_2 = 0.5W$ \emph{Inverse
% Channel}
% }{uni_inverse}
%
% \begin{figure*}[!t]
% \centering
% \subfloat[]{\includegraphics[width=2.5in]{graphics/1}
% }
% \hfil
% \subfloat[]{\includegraphics[width=2.5in]{graphics/2}
% }
% \caption{$\alpha_1 = 0.25,f_1 = 0.25W, \alpha_2 = 0.25 , f_2 =
% 0.75W$
% }
% \label{fig:multi}
% \end{figure*}

% \nfig{graphics/1}{graphics/2}{$\alpha_1 = 0.25,f_1 = 0.25W, \alpha_2 = 0.25 , f_2 =
% 0.75W$
% }{multi}
It is clear from the results that:
\begin{itemize}
\item The proposed approach for allocating the power $S(f)$ and rate $C(f)$ is indeed optimal and
superior to the other methods that were presented. Evidently, the rate is
upper bounded by the classical water-pouring result~($C\rightarrow\infty$). It is
evident that
\begin{align}
I_n^\infty(X;Z) \ge& I_n^C(X;Z)\\ \nonumber\ge& I_n^C(X;Z)|_{\text{Limited-Rate Water-Pouring}}\\ \nonumber \ge& I_n^C(X;Z)|_{\text{Uniform Allocation}}
\end{align}
\item The price of obliviousness is demonstrated; as for a cognitive
relay the reliable rate is $min(I_n^\infty(X;Z),C)$,
achieved by a relay that decodes the signal and then transmits
the decoded information at the maximum allowable rate ($C$).
\end{itemize}

%% file: finite_entropy.tex
%\chapter{Finite Output Entropy}
\section{Finite Output Entropy $H(Z)$}
\label{sec:finite_entropy}
In this section we analyze the performance of finite output entropy quantizers that can be implemented by a standard Lempel-Ziv algorithm, at a small cost in terms of performance. Analytic solutions for optimal information bottleneck quantizers are rarely
available; here, we investigate optimization algorithms, since most practical algorithms cannot guarantee reaching a global optimum \cite{s.hassanpourd.wuebbena.dekorsy2017}.
% \nfigSing{graphics/ixz_hz_leq_c_2bit}
%  {$I(X;Z)$ vs SNR - 2 Bit Quantizer}
%  {fig:IVWvsSNR}
 
% \nfigSing{graphics/hz_vs_snr_2bit}{$H(Z)$ vs SNR - 2 Bit Quantizer}{fig:HVvsSNR}
\subsection{Quantizer Model and Preliminaries}
Reviewing the scalar bottleneck problem, we assume 
\beqd{\label{eq:finite_entropy_channel_model}
Y = \sqrt{snr}\cdot X +N,} 
where $X$ and $N$ are unit variance independent Gaussian signals, and, hence, $Y$ designates the output of a scalar Gaussian channel with a Gaussian input. The Finite-Entropy-Bottleneck Problem, reads: Find the maximum of $I(X;Z)$ under the Markov condition $X - Y - Z$, where $H(Z) = C$. 
In mathematical form:
\beqdn{\max_{P_{Z|Y}(Z|Y) : H(Z) = C} I(X;Z).}
As mentioned in the preliminaries, the
deterministic solution is optimal. In order to make computation feasible, the search was carried out for a K-bin or (K-level) deterministic quantizer $\hat{Q}$. $\hat{Q}$ maps the real input $Y$ into one of K-bins, $Z = \hat{Q}(Y)$, producing discrete outputs with alphabet $Z \in \chi ,  |\chi| < \infty $. Bear in mind that $H(Z) \le C$, but now $H(Z|Y) = 0$ since the quantizer is deterministic: hence $I(Y;Z) = H(Z)$.\newline 
First we list a few definitions: \newline
Assume that $Z = z_i$ if $Y \in [q_{i-1},q_i]$. We know that $P_{Y|X}(y|x) = N\big(\sqrt{snr}\cdot x,1\big), \sigma_Y  = \sqrt{1+snr} $ and hence the probabilities $P_{Z|X}(z|x)$ and $P_Z(z)$ are 
\begin{align}
P_{Z|X}(z_i|x) =&\;p_{Y|X}(q_{i-1}\leq y \leq q_i|x) \nonumber\\  = &\; Q\Big(q_{i-1}-\sqrt{snr}\cdot x\Big) - Q\Big(q_{i}-\sqrt{snr}\cdot x\Big),  \nonumber \\
P_Z(z_i) =&\;p_{Y}(q_{i-1}\leq y \leq q_i) \nonumber\\  = &\; Q\Big( \frac{q_{i-1}}{\sigma _Y}\Big) - Q\Big( \frac{q_{i}}{\sigma _Y}\Big). \nonumber 
\end{align}
The resulting $H(Z|X),H(Z)$ and $I(X;Z)$ are
\begin{align*}
&H(Z|X=x) =  \sum _{z_i \in \chi} \Big[ -P_{Z|X}(z_i|x) \cdot \log\Big(P_{Z|X}(z_i|x)\Big) \Big], \nonumber \\
&H(Z|X) \quad \;\;\,=  \int dF(x) H(Z|X=x), \nonumber \\
&H(Z) \quad \quad \;\;\,\,= \sum_{z_i\in \chi}\Big[-P_Z(z_i)\log P_Z(z_i)\Big], \nonumber \\  
&I(X;Z) \quad \;\;\;=  H(Z) - H(Z|X).
\end{align*}
$Q(x)$ denotes the complementary Gaussian distribution function $\frac{1}{\sqrt{2\pi}}\int_x^\infty e^{-t^2/2} dt$ . $F(x)$ is the cumulative distribution function (cdf) of $x$. \newline \newline
Since both the information source and the noise are symmetric, we limit ourselves to the class of symmetric quantizers such as Eq.~(\ref{eq:symmetric_quan_def}). 
%\hl{As a side-note, I wasn't able to prove that our problem is convex in the set of quantizers thresholds $\{ q_i\}$ and as a result, I couldn't use the cutting plane algorithm and had to use brute force optimization}
The optimal quantizer problem can be stated as follows:
\beqdn{%\label{eq:optimal quantizer}
\max_{\{q_i\} : H(Z)\le C} I(X;Z).}
The maximization is performed over the quantizer thresholds, $\{q_i\}$. In the following subsections we present numerical results of the problem under various conditions, and will gain some insights on the nature of the optimal quantizer and develop bounds and an analytical approximation.

\subsection{Numerical Analysis}

% \nfigSing{graphics/ixz_hz_leq_c_3bit}
%  {$I(X;Z)$ vs SNR , 3-Bit Quantizer}
%  {fig:ivwsnr3bit}

% \nfigSing{graphics/hz_vs_snr_3bit}
% {$H(Z)$ vs SNR , 3-Bit Quantizer}
% {fig:hvsnr3bit}

% We start with the 2-bit symmetric quantizer with thresholds $\{-q,0,q\}$, we've calculated the resulted $I(X;Z), H(Z)$ for various types of SNR and $C$ as can be seen by Figs. ~\ref{fig:IVWvsSNR},~\ref{fig:HVvsSNR} respectively.
Our numerical optimization yields a 3-bit symmetric quantizer with thresholds
\beqd{\{q_i\}_{1}^{7} = \{-q_3,-q_2,-q_1,0,q_1,q_2,q_3\}.\label{eq:symmetric_quan_def}}
The thresholds were optimized to maximize the mutual information $I(X;Z)$ for various types of SNR and $C$. In Fig.~\ref{fig:ULBoundsPMF} we see the resulting mutual information, as well as upper and lower bounds.
From the results we see that:
\begin{itemize}
\item The mutual information $I(X;Z)$ increases with SNR and $C$
%\item The output entropy is bounded by $\log_2 |\chi|$ as expected
% \item The simulation result shows that the quantizers thresholds become distinguishable at high SNR whereas at low SNR they tend to "encapsulate" the entire signal (bear in mind that as the SNR increase the power of the measurement $Y$ increases) 
\item The mutual information is bounded by the GIB.
\end{itemize}
% We now examine the influence of bit-rate constraint at the relay, $C$, on the total mutual information $I(X,Z)$.
%In particular, it is interesting to see the performance when $C\ge \log_2 |\chi|$ and $C\le \log_2 |\chi|$ in the case when we limit the quantizer output $z$ to some alphabet $\chi$.
\subsection{The Effect of an Entropy Constraint on the Deterministic Quantizer Operation}
We examine the case of an entropy constraint deterministic quantizer ($C \le \log_2 |\chi|$ , when $Z\in\chi$). From Fig.~\ref{fig:ivw_qcomparison_c2} it is evident that increasing the number of levels of the quantizer above the entropy constraint has almost no effect on the mutual information; thus the number of bins used was sufficient. The mutual information is bounded, as expected, by $\log_2 |\chi|$. One can see that even in a low SNR scenario the difference between the quantizers is negligible. To complete the discussion we add the case of a memoryless deterministic quantizer (i.e. no constraint, $C \geq \log_2 |\chi|$), as illustrated in Fig.~\ref{fig:CgeqQ}. Here, unlike the previous cases, there is a clear gain using a quantizer which has more bits.

\subsection{Lower and Upper Bounds on the Optimal Performance}
We now try to bound the mutual information and apply an upper bound and two lower bounds. 
As before, the GIB can serve as an upper bound.
For the lower bound (which are also interesting achievability schemes), we tested two schemes: 
\subsubsection{Lower bound - setting output entropy $H(Z) = C$}
 \nfigSing{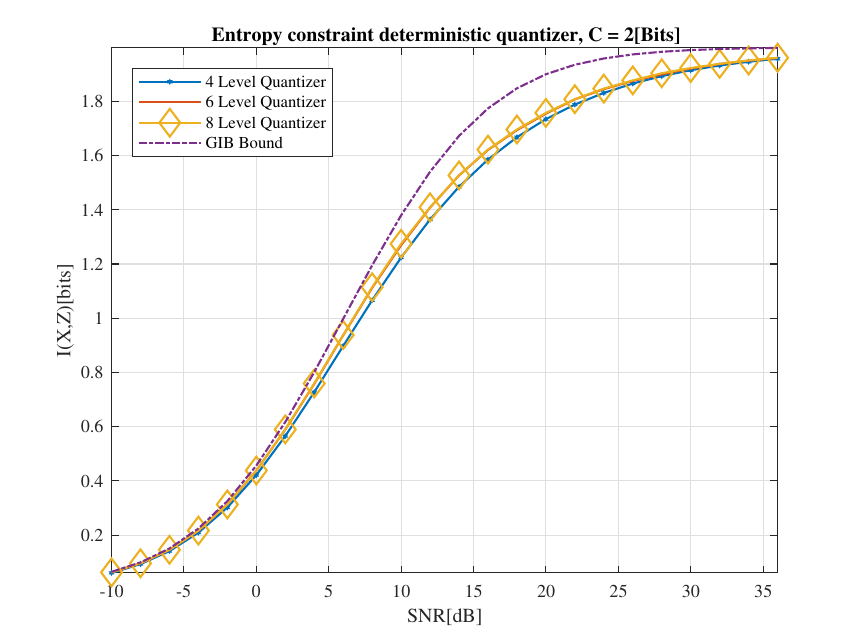}
 {Entropy constraint deterministic quantizer $C = 2[bits]$. The performance with six and eight levels is nearly identical.}{fig:ivw_qcomparison_c2}
 \nfigSing{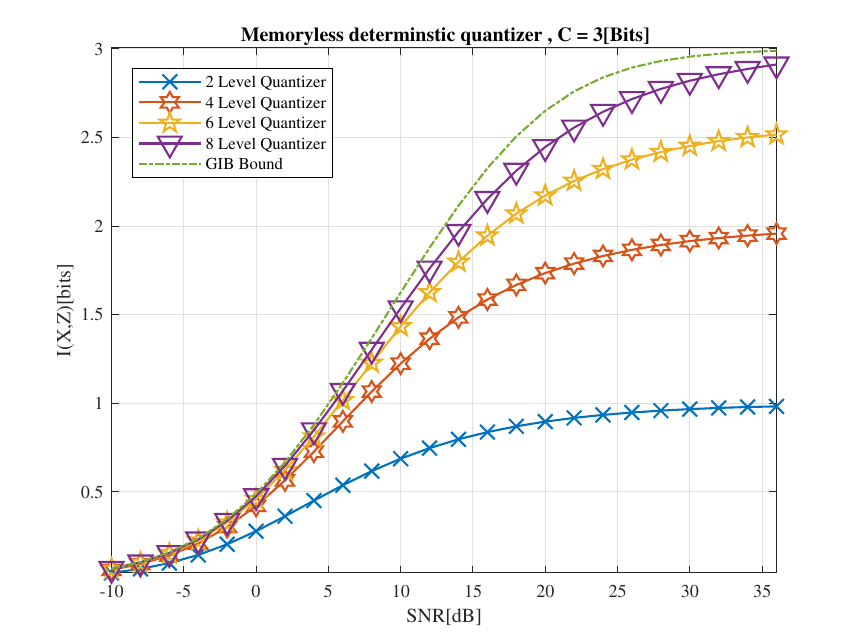}{Memoryless deterministic quantizer, $C = 3[bits]$.}{fig:CgeqQ}
We chose a quantization scheme which will lead to an output entropy $H(Z) = C$.
In order to assure the required entropy, we changed the cardinality of the output $|\mathcal{Z}|$ and the induced (probability mass function) $P_Z(z)$ using the method described in Sec.~\ref{alg:change_induced_pmf}. Once the output probability mass function $P_Z(z)$ was set, the (symmetric) quantizer thresholds, $\{q_i\}_{1}^{|\mathcal{Z}|-1}$, can be found by taking an auxiliary variable $\nu_i$:
\beqdn{\nu_i = \sum_{z=1}^i P_Z(z).} 
The threshold $q_i$ is
\beqd{q_i = \sigma_Y \cdot Q^{-1}(\nu_i),}
where $Q^{-1}(x)$ denotes the inverse of $Q(x)$. 
Fig. \ref{fig:ULBoundsPMF} demonstrates these results.
 
 \nfigSing{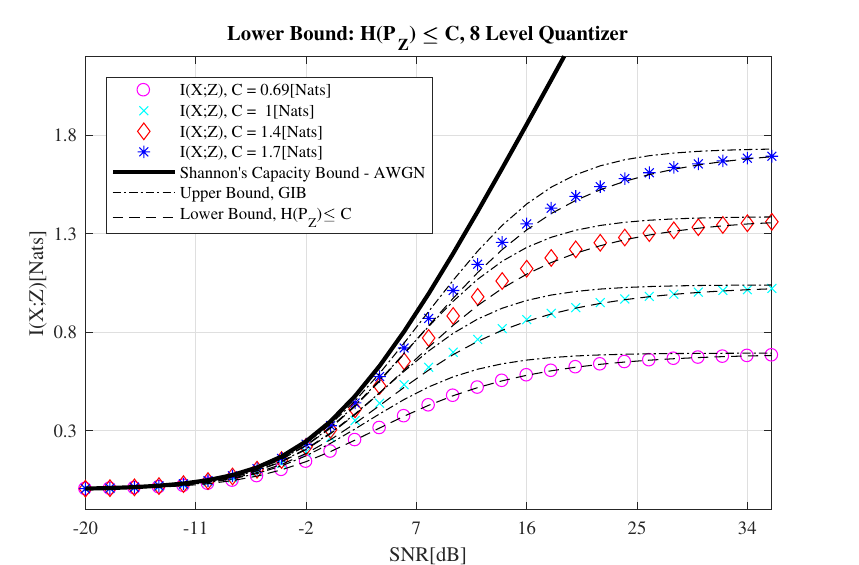}{Lower bound: setting the probability mass function $P_Z(z)$ s.t $H(Z)\leq C$, for each bit-rate constraint $C$, we present the numerical result for the optimal quantizer, the GIB upper bound, and the lower bound resulting from setting $H(P_Z)\leq C$.}
 {fig:ULBoundsPMF}
 
\subsubsection{Lower bound - uniform quantizer}
We tested a uniform quantizer, in which the quantizer step $q$ was increased until the resulting probability mass function $P_Z(z)$ of the quantizer output had output entropy $H(P_Z) = C$. The output of the uniform quantizer has infinite cardinality since its input is unbounded. To that end, we discarded values that are higher (in their absolute value) than $N\cdot\sigma_Y$, ensuring output cardinality of $|\mathcal{Z}| \approx \frac{2N\sigma_Y}{q}$ for some large $N$.

Fig.~\ref{fig:ULBoundsUniQ} presents the results. For each bit-rate constraint $C$, we plot the numerically optimized quantizer, the GIB upper bound, and the lower bound resulting from uniform quantization. As one can see, the lower bound is fairly near to the curve of the numerically optimized quantizer. This method produced a tighter bound than the previous.

\nfigSing{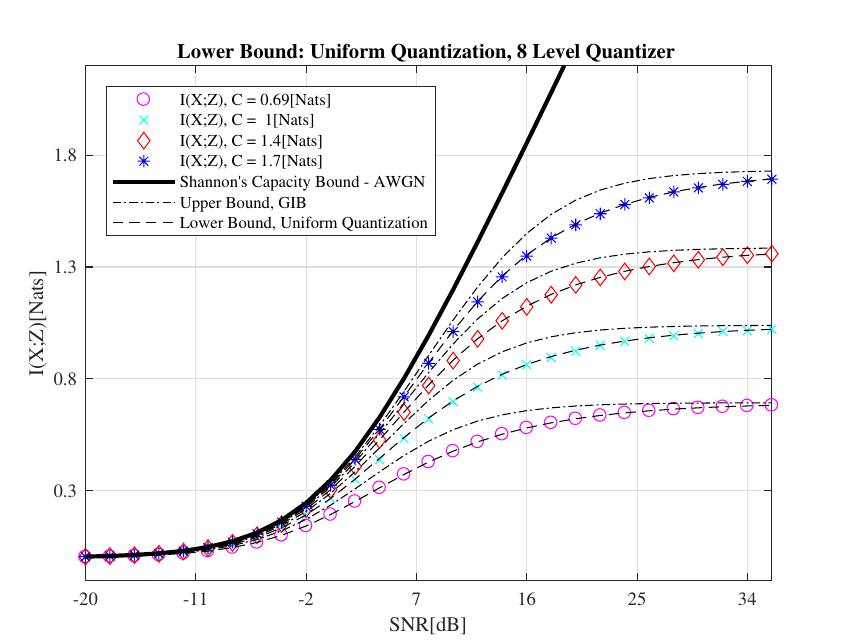}{Lower bound: Uniform quantization. For each bit-rate constraint $C$, we present the numerical result for the optimal quantizer, the GIB upper bound, and the lower bound resulting from uniform quantization of the channel output.}{fig:ULBoundsUniQ}
\nfigSing{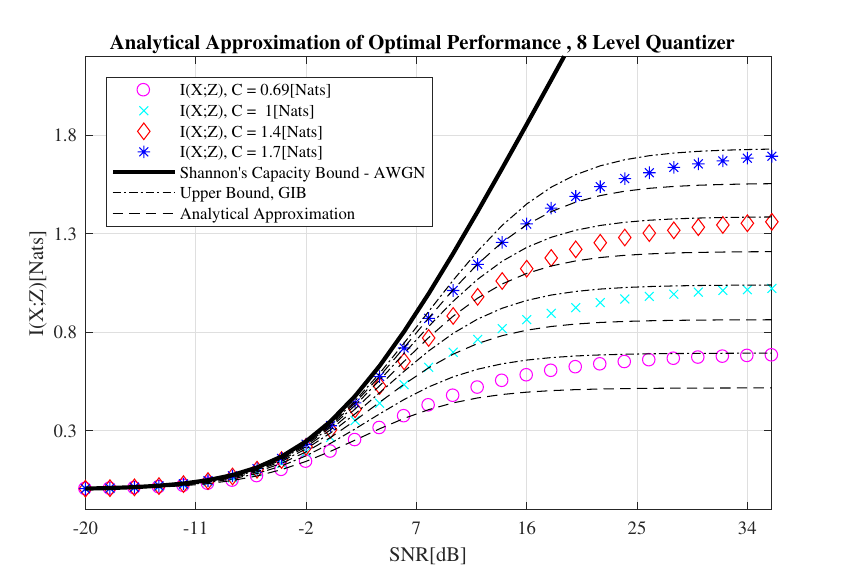}{Analytical approximation: For each bit-rate constraint $C$, we present the numerical result for the optimal quantizer, the GIB upper bound, and the analytical approximation.}
 {fig:AnalyticLB}
 
\subsection{Analytic approximation of optimal performance}
Let $Z - Y - X$ be the inverse of the Markov chain defined in Sec.~\ref{sec:finite_entropy} and Eq.~(\ref{eq:finite_entropy_channel_model}).
Define $Y - X$, the inverse channel, as
\beqdn{X = E[X|Y] + (X - E[X|Y]).}
Thus, $X$ can also be written as
\beqdn{X = \frac{\sqrt{snr}}{1+snr} Y + \frac{1}{\sqrt{1+snr}}M.}
Since $E[X|Y] = \alpha Y$ (where $\alpha = \frac{\sqrt{snr}}{1+snr}$), and due to the fact that the error term $X- E[X|Y]$ is independent of the measurement $Y$, $M$ is a normalized Gaussian variable independent of $Y$. Having done so, note that
\begin{align}
I(X;Y,Z) = & I(X;Z) + I(X,Y|Z) \nonumber \\ = &  I(X;Y)  + I(X;Z|Y) \nonumber \\ = & I(X;Y)  + 0 .\nonumber
\end{align}
$I(X;Z|Y) = 0$ due to Markovity, leading to
\beqdn{I(X;Z) = I(X;Y) - I(X,Y|Z).}
Then $I(X;Y|Z)$ is no more than a standard
Gaussian channel from $Y \rightarrow X$, but $Y$ is conditioned on $Z$ since Gaussian inputs are optimal given the variance constraint
\beqdn{
\label{eq:ixygz_gaus_input}
I(X;Y|Z) \le E_Z\Bigg\{\frac{1}{2} \log \Big(\frac{snr}{1+snr}VAR(Y|Z) + 1\Big)\Bigg\},} 
where $VAR(Y|Z) = E_{Y|Z}\{[Y-E(Y|Z)]^2|Z\}$.
Incorporating the Jensen inequality will lead to
\beqd{
\label{eq:upper_bound_ixy|z}
I(X;Y|Z) \le \frac{1}{2} \log \Big(\frac{snr}{1+snr}MMSE(Y|Z) + 1\Big),} 
where the $MMSE(Y|Z) = E[Y-E(Y|Z)]^2$ is the MMSE error of $Y$ given $Z$. At this point we can utilize the results of Gish~\cite{h.gishj.n.pierce1968}, where $E[Y-E(Y|Z)]^2$ is minimized under the constraint
of the entropy of $Z, H(Z)$, leading to the lower bound
\begin{align}
\label{eq:mui_lb}
I(X;Z) \ge & \frac{1}{2} \log{(1+snr)} \:- \\ & \frac{1}{2} \log \Big(\frac{snr}{1+snr}MMSE(Y|Z) + 1\Big) .\nonumber
\end{align}
Now, from Gish (when large output entropy is permitted or the quantization interval tends to zero),
\beqdn{C[bits] = H(Z) \approx 0.5\log_2 \Big(\frac{\sigma^2_Y}{MMSE(Y|Z)}\Big) + 0.255[bits],}
where $\sigma^2_Y = 1 +snr$; hence (converting bits to nats),
\beqd{\label{eq:gish_mmse} MMSE(Y|Z) \approx (1+snr)e^{(0.354-2C)}.}
One can see that the quantization noise ,$MMSE(Y|Z)$ decreases as $C \rightarrow \infty$, and increases with SNR (as the power of $Y$ increases with it). 
As mentioned by Gish, the approximation is tight for low quantization noise and high output entropy (i.e. both  $MMSE(Y|Z)$ and the SNR tend to $0, C\rightarrow\infty$).
% , since, as the SNR increases, the $MMSE(Y|Z)$ increase as well and the approximation for the output entropy $H(Z)$ is no longer valid.
Incorporating (\ref{eq:gish_mmse}) into (\ref{eq:mui_lb}) will lead to
\beqdn{
I(X;Z) \ge \frac{1}{2} \log{\Big(\frac{1+snr}{1+snr\cdot e^{(0.354-2C)}}\Big).}
}

Massey \cite{jamesmassey1976} has proved that in an AWGN channel at low SNR and with a zero-mean input, the capacity is the same function of the
mean power regardless of the input's probability distribution function.
It is also evident that zeroing the added component $0.354[nats]$ leads to the GIB and Gish's bounds coinciding, since
\beqd{\label{eq:accurate_mmse} MMSE(Y|Z) = (1+snr)e^{-2C}.}
Incorporating (\ref{eq:accurate_mmse}) in (\ref{eq:upper_bound_ixy|z}) will lead exactly to the GIB bound, which is achieved in the case where the inverse channel input $Y|Z$ is Gaussian, as GIB dictates. Fig.~\ref{fig:AnalyticLB} demonstrates these results. 
Thus, the difference in performance at a low SNR and high $C$ between the
stochastic mutual information constrained quantizer and the deterministic entropy
constrained quantizer is exactly the 0.255 bits per symbol in the relay bit-rate $C$.

%% file: details_and_proves.tex
\section{Further Derivations and Proofs}
\label{sec:proves}

\subsection{Complete Derivation of Solution of Eq.~(\ref{eq:lagrange})}
\label{sec:deriv_of_two_sltns}
Differentiating Eq. (\ref{eq:lagrange}) with respect to $\hat{C},\hat{S}$ leads to
\begin{subequations} \label{eq:diffL}  
\begin{align} 
   \dd{I[f,\hat{S},\hat{C}]}{\hat{S}}- \lambda _s & = 0, \label{eq:diffLa} \\
   \dd{I[f,\hat{S},\hat{C}]}{\hat{C}} - \lambda _c & = 0. \label{eq:diffLb}
\end{align}
\end{subequations}
Hence,
\begin{subequations} 
\begin{align} 
   0 &= \frac{H(f)^2(1-e^{-\hat{C}})}{(1+\hat{S} H(f)^2)(1+\hat{S}H(f)^2e^{-\hat{C}})} -\lambda_s,  \label{eq:S}\\
    0 &= \frac{\hat{S} H(f)^2e^{-\hat{C}}}{1+\hat{S}H(f)^2e^{-\hat{C}}} - \lambda _c.\label{eq:Q}
\end{align}
\end{subequations}
In order to simplify notation we use the following definitions:
\beqdn{\hat{Q} \triangleq \exp(-\hat{C}),}
\beqdn{X_f \triangleq \Xf.}
From (\ref{eq:Q}) it is clear that
\beqd{\label{eq:SfromQ}
\hat{S} = \frac{\lambda_c}{\hat{Q}H(f)^2(1-\lambda_c)}.}
Substituting (\ref{eq:SfromQ}) in (\ref{eq:S}) will lead to Eq. (\ref{eq:quadQ}). We now have two sets of solutions for $\{\hat{S},\hat{Q}\}$.

Define
\beqdn{
\psi_i \equiv 
\begin{cases}
1 \quad &i=1 \\
-1 \quad &i=2 
\end{cases}
.
}
Then the solution for $\{\hat{S},\hat{Q}\}$ is
\begin{subequations}
\label{eq:solLtmp}
\begin{align}
   &\hat{S}_{i}=\frac{2\lambda_c}{X_f-\psi_i \sqrt{X_f^2-4H(f)^2\lambda _c \lambda _s}},     
   \\    
   &\hat{Q}_i=\frac{X_f-\psi_i \sqrt{X_f^2-4H(f)^2\lambda _c \lambda _s}}{2H(f)^2(1-\lambda _c)}.   
\end{align}
\end{subequations}
Multiplying the denominator and numerator by
\beqdn{X_f+\psi_i \sqrt{X_f^2-4H(f)^2\lambda _c \lambda _s}}
will lead to
\begin{subequations}
\label{eq:solL1}
\begin{align}
   \hat{S}_{i}=\frac{X_f+\psi_i\sqrt{X_f^2-4H(f)^2\lambda _c \lambda _s}}{2H(f)^2\lambda _s},   
   \\    
   \hat{Q}_{i}=\frac{X_f-\psi_i \sqrt{X_f^2-4H(f)^2\lambda _c \lambda _s}}{2H(f)^2(1-\lambda _c)}.   
\end{align}
\end{subequations}
At this point, we continue in accordance with Proposition~\ref{prop:take_concave_sol} and discard the $\{\hat{S}_{2},\hat{Q}_{2}\}$ curve since it is a non-concave solution.
\subsection{Proof of Proposition~\ref{prop:bound_lambda}}
\label{subsec:bound_lemma}
By investigating the derivatives of $I[f,\hat{S},\hat{C}]$ w.r.t $\{\hat{S},\hat{C}\}$ and taking into account that $\hat{S} \ge 0,\hat{C} \ge 0$ one can see that 
\begin{subequations} \label{eq:bdbound} 
\begin{align}
   \lambda _s & = \dd{I[f,\hat{S},\hat{C}]}{\hat{S}} \\  & = \frac{H(f)^2(1-e^{-\hat{C}})}{(1+\hat{S}H(f)^2e^{-\hat{C}})(1+\hat{S}H(f)^2)}
   > 0,    \nonumber \\
   \lambda _s & = \dd{I[f,\hat{S},\hat{C}]}{\hat{S}}  \le H(f)^2 \le \max{H(f)^2},\\
   \lambda _c & = \dd{I[f,\hat{S},\hat{C}]}{\hat{C}}  =\frac{\hat{S}H(f)^2e^{-\hat{C}}}{(1+\hat{S}H(f)^2e^{-\hat{C}})} \ge 0, \\
   \lambda _c & = \dd{I[f,\hat{S},\hat{C}]}{\hat{C}}  = \frac{\hat{S}H(f)^2e^{-\hat{C}}}{(1+\hat{S}H(f)^2e^{-\hat{C}})} \le
   1.   
\end{align}
\end{subequations}
The bounds for $\{\lambda _c,\lambda _s\}$ follow from (\ref{eq:bdbound}).
\subsection{Constructing the Set of Operating Frequencies $\mathcal{B}_l$}
\label{alg:setting_bl} 
We perform a bounded grid search (see Proposition~\ref{prop:bound_lambda}) on $\{\lambda_s,\lambda _c\}$ that will yield the maximum mutual information:
\beqdn{\intw{I[f,S(f),Q(f)]}.} 
For each $\lambda_s,\lambda_c$, the produced curves of $S(f),Q(f)$ (and hence, $S(f),C(f)$) might not meet the resource constraint ($\intw{S(f)}>P$ or $\intw{C(f)}>C$). At this point, we sort $I[f,S(f),Q(f)]$ and discard the frequencies (i.e. $S(f) = 0 ,Q(f) = 1$) that contribute least to the total mutual information, until compliance. The set of frequencies that were not discarded is $\mathcal{B}_l$.

%% file: conclusions.tex
\section{Conclusions and Future Directions}
\label{sec:conclusions}
We presented and analyzed the rate- and power-limited oblivious
relay over the frequency selective AWGN channel and derived
the optimal transmit power spectral density and the optimal
allocation of the relay bit-rate for Gaussian signaling. Our results relate directly to the
classical water-pouring method, as well as to the Gaussian bottleneck
frameworks. The advantage of this approach over other methods
was demonstrated. We also investigated the class of finite entropy quantizers and, while it is difficult to find an analytical expression for the optimal quantizer, we devised lower and upper bounds for this case.

Our results on water-pouring also apply directly to the frequency dependent vector (MIMO) channels. Such channels can be transformed to a set
of parallel independent channels \cite{l.brandenburga.wyner1974}. Thus, equation (\ref{eq:solL1}) and the optimization algorithms can be
applied on them with no need for modification by considering those independent
channels as occupying independent frequency bands. A modern implementation of such a MIMO system
might use the (Orthogonal Frequency-Division Multiplexing) OFDM framework in which the MIMO channel diagonalization is
convenient to implement (see, for example, \cite{j.mietznerr.schoberl.lampew.h.gerstackerp.a.hoeher2009}).
One could extend the method presented in this paper to the setting where only partial channel state information is available
to the transmitter. For example, the transmitter may assume a flat channel and a lower bound on
the SNR at the relay. We expect such an approach to improve the performance with respect to other methods since the optimal
scheme would use only part of the spectrum, as presented above.

%% file: appendix.tex
\section{Appendix}
\subsection{Proof of equivalence between entropy constraint stochastic and deterministic quantizers }
\label{sec:equivalnce_stochastic_deterministic_quantizers}
A stochastic quantizer with limited $H(z)$ over the AWGN channel is characterized by $P_{Z|Y}(z|y)$ . We
construct a deterministic quantizer with the same performance as follows.
Divide the range of $y$ into segments $\gamma_j$ small enough so that in each segment $p_{Y|X}(y|x)$ changes as
little as desired. Denote by $y_j$ the value of $y$ in the center of $\gamma_j$. Divide each segment $\gamma_j$ into subsegments,
each mapped into a different $z_k$ by the deterministic quantizer so that $P_{Z|Y}(z_k|\gamma_j)$ is preserved. The division is straightforward since in each segment $p_{Y|X}(y|x)$ is as constant as desired
for all $x$ and, clearly, so is $p_Y(y)$. Then, the probability of each subsegment when $\gamma_j$ is given is the
ratio of the length of the subsegment to the length of the segment. Thus, in each segment, each
$z_k$ is mapped to a subsegment the length of which is proportional to $P_{Z|Y}(z_k|\gamma_j)$ in the original
stochastic quantizer. 
To prove equal performance of both the quantizers it is sufficient to
establish that $P_{Z|X}(z_k|x)$ is preserved since it determines $I(x;z)$ , $P_Z(z)$ and $H(z)$.
For the stochastic quantizer and using the relations $p_Y(y)= p_Y(y_j)$, $p_{Y|X}(y|x) = p_{Y|X}(y_j|x)$ holding to any
desired accuracy in each segment, we have
\begin{subequations}
\begin{align}
P_{Z|X}(z_k|x) &=\int_{-\infty}^{\infty}P_{Z|Y}(z_k|y)p_{Y|X}(y|x)dy \label{eq:use_markovity}
\\
&=\sum _j\int_{\gamma_j}P_{Z|Y}(z_k|y)p_{Y|X}(y|x)dy
\\
&=\sum _j\int_{\gamma_j}\frac{P_{Z|Y}(z_k,y)}{p_{Y}(y)}p_{Y|X}(y|x)dy
\\
&=\sum _j\frac{p_{Y|X}(y_j|x)}{p_{Y}(y_j)}\int_{\gamma_j}p_{Z,Y}(z_k,y)dy
\\
&=\sum _j\frac{p_{Y|X}(y_j|x)}{p_{Y}(y_j)}p_{Z,Y}(z_k,\gamma_j)
\\
&=\sum _j\frac{p_{Y|X}(\gamma_j|x)}{p_{Y}(\gamma_j)}p_{Z,Y}(z_k,\gamma_j)
\\
&=\sum _jp_{Y|X}(\gamma_j|x)P_{Z|Y}(z_k|\gamma_j). \label{eq:dtr_stc_equiv}
\end{align}
\end{subequations}
Where (\ref{eq:use_markovity}) stems from Markovity. (\ref{eq:dtr_stc_equiv})  clearly also holds for the deterministic quantizer. 
\subsection{Proof of Proposition \ref{prop:take_concave_sol}}
\label{apndx:proof_of_concave_sign}
In this subsection we prove that we can discard the solution  $\{S_{2}(f),Q_2(f)\}$ of (\ref{eq:solL1}), based on the concavity of I[f,S(f),C(f)] on the set $\{S(f),C(f) \}$ at each fixed frequency $f$ (for simplicity we discard the channel dependence from now on and denote $I[S(f),C(f)]$).  
To this end, we shall prove that:
\begin{itemize}
\item Any point in the optimal solution, cannot reside in the non-concave region of $I[S(f),C(f)]$. 
\item The regions of concavity of $I[S(f),C(f)]$ coincide with the $\{S_{1}(f),C_1(f)\}$ solution.    
\end{itemize}

The optimal solution at any frequency $f$ cannot be in the non-concave region of $I[S(f),C(f)]$, because
such a solution can be improved as follows: Suppose the solution assigned the resources $df\cdot S$ and $df\cdot C$ in an infinitesimal frequency band $df$ around $f$ such that $I[S(f),C(f)]$ is not concave at this point. Then, the $df$ band can be split into two sub-bands with the resource assignment
perturbed in each, but with the sum of the resources in $df$ unchanged while increasing the performance $I$ in $df$ using the non-concavity. This is to be expected since our optimization equations are necessary but not sufficient conditions of global optimality~\cite{bib.gelfand}. 
% Let us assume the \textit{optimal} solution has points that lies in the non-concave region of $I[f,S(f),C(f)]$, We could a hair splitting operation $I[f,S(f),C(f)],I[f,S(f)+\Delta_s,C(f) + \Delta_c]$, and achieve a better rate, which render the solution non optimal.   

Since we are dealing with a single frequency, $H(f)$ is constant and its influence is only a scaling of $S(f)$. Let us rewrite the Lagrangian at (\ref{eq:lagrange}):
\beqd{
L = log\bigg(\frac{1+S}{1+Se^{-C}}\bigg)-\lambda_s S -\lambda_c C  
}
and equation (\ref{eq:solL1}) becomes (remembering that $Q\triangleq e^{-C} , C = -log(Q)$)
\begin{subequations}
\label{eq:discard_f}
\begin{align}
   &S  =  \frac{1-\lambda _c - \lambda _s  +\psi_i\sqrt{(1- \lambda _c - \lambda _s)^2-4\lambda _c \lambda _s}}{2\lambda _s},   
   \\    
   &Q  = 
    \frac{1-\lambda _c - \lambda _s -\psi_i\sqrt{(1- \lambda _c - \lambda _s)^2-4\lambda _c \lambda _s}}{2(1-\lambda _c)}.   
 \end{align}
\end{subequations}
We can also write $\lambda_s,\lambda_c$ as a function of $(S,Q)$:
\begin{subequations}
\label{eq:lambda_sc_func_SQ}
\begin{align}
   &\lambda _c = -\frac{\partial L}{\partial C} =\frac{SQ}{1+SQ},    
   \\    
   &\lambda _s = -\frac{\partial L}{\partial S} =\frac{1}{1+S} - \frac{\lambda _c}{S}.   
\end{align}
\end{subequations}
We would like to choose only the concave solution, that is to choose $(S,C)$ such that
\begin{subequations}
\label{eq:concave_def}
\begin{align}
\frac{\partial^2 I}{\partial S^2}\frac{\partial^2 I}{\partial C^2} - \frac{\partial^2 I}{\partial S \partial C}\frac{\partial^2 I}{\partial C \partial S} \geq 0 ,\\
\frac{\partial^2 I}{\partial S^2} \leq 0 \qquad and \qquad \frac{\partial^2 I}{\partial C^2} \leq 0. 
\end{align}
\end{subequations}
We then prove that regions with a value of $\Psi _i = 1$ and concavity are identical. 
\begin{lem}
\label{lemma:psi}
Different $\psi_i$ in (\ref{eq:discard_f}) enforce different $(S,C)$.
\end{lem}

\begin{IEEEproof}
Each $(S,C)$ pair corresponds by (\ref{eq:lambda_sc_func_SQ}) to a unique $(\lambda _s,\lambda _c)$.  Thus, the same $(S,C)$ cannot be the outcome of two distinct $(\lambda _s,\lambda _c)$ with different $\psi$.
\end{IEEEproof}

The next step will be to show that the lines $S(\lambda _s,\lambda _c) = f(C;\lambda _s,\lambda _c)$ that split the regions of concavity and the sign of $\psi$ coincide. 
Let us derive the dividing line between the $\pm$ regions in (\ref{eq:discard_f}). At the dividing line $(1+ \lambda _c + \lambda _s)^2-4\lambda _c \lambda _s$ must be zero by the proof of Lemma~\ref{lemma:psi} and the fact the functions are continuous, so at any point the dividing line must be the result of (\ref{eq:discard_f}) regardless of the value of $\psi$. Particularly:
two points infinitesimally near and each on a different side of the dividing line have
the same $S,C$ in the limit and, on the other hand, also the same $(\lambda _s,\lambda _c)$ by (\ref{eq:lambda_sc_func_SQ}), so in the limit the value of $\psi$ will not matter.
\beqd{
\label{eq:discriminant_zero}
(1- \lambda _c - \lambda _s)^2-4\lambda _c \lambda _s = 0,}
leading to
\beqd{
\label{eq:lambdas_sol}
\lambda_{s,i} = (\lambda_c + 1) + \eta _i \sqrt{4\lambda _c} 
,}
where
\beqd{
\eta_i = 
\begin{cases}
1 \quad &i=1 \\
-1 \quad &i=2 
\end{cases}
.
}
Substituting  (\ref{eq:lambdas_sol}) back into (\ref{eq:discard_f}) yields

\beqd{S_i = \frac{1-\lambda _{s,i} - \lambda _c }{2\lambda _{s,i}}=\frac{\lambda _c + \eta _i \sqrt{\lambda _c}}{\lambda _c + 1 + 2\eta _i \sqrt{\lambda _c}}.}

The allocated power $S_i$ must be non-negative; hence, by elimination, we discard $\eta _1$:
\beqd{
\label{eq:S_sol_sign}
S_{plus/minus}  = 
\frac{-(\lambda _c - \sqrt{\lambda _c})}{\lambda _c + 1 -2 \sqrt{\lambda _c}} = \frac{\sqrt{\lambda _c}}{1-\sqrt{\lambda _c}},
}
where $S_{plus/minus}$ is $S$ on the dividing line defined by the sign of $\psi _i$. 
Remembering that $Q = \frac{1}{S}\frac{\lambda _c}{1-\lambda _c}$ and $C=-log(Q)$ we have the curve $(S(\lambda _c),C(\lambda _c))$. % which is exactly the line at fig. \ref{fig:plot_concave_psi}.
We now examine the concavity regions of (\ref{eq:concave_def}). 
We use the following derivatives:
\begin{subequations}
\label{eq:deriv_res}
\begin{align}
\frac{\partial^2 I}{\partial S^2} &= -\frac{1}{(1+S)^2} + \frac{e^{-2C}}{(1+Se^{-C})^2}  \nonumber \\ &=  -\frac{1}{(1+S)^2} + \frac{\lambda _c}{S^2},\\
\frac{\partial^2 I}{\partial C^2} &= \frac{S^2e^{-2C}}{(1+Se^{-C})^2} - \frac{Se^{-C}}{1+Se^{-C}} = \lambda ^2 _c - \lambda _c,\\
\frac{\partial^2 I}{\partial S\partial C} &=
\frac{\partial^2 I}{\partial C\partial S}  \nonumber \\ &=-\frac{Se^{-2C}}{(1+Se^{-C})^2} + \frac{e^{-C}}{1+Se^{-C}} \nonumber \\ &= \frac{1}{S}(\lambda ^2 _c - \lambda _c).
\end{align}
\end{subequations}
 It is easy to see that $\frac{\partial^2 I}{\partial S^2} < 0 $ and $\frac{\partial^2 I}{\partial C^2} < 0$, but we need to examine the sign regions of $\frac{\partial^2 I}{\partial S^2}\frac{\partial^2 I}{\partial C^2} - \frac{\partial^2 I}{\partial S \partial C}\frac{\partial^2 I}{\partial C \partial S}$.
Substituting (\ref{eq:deriv_res}) in (\ref{eq:concave_def}) we get
\begin{align}
\label{eq:quad_S_concave}
0 & = \frac{\partial^2 I}{\partial S^2}\frac{\partial^2 I}{\partial C^2} - \frac{\partial^2 I}{\partial S \partial C}\frac{\partial^2 I}{\partial C \partial S} \nonumber \\
& = \Big[ -\frac{1}{(1+S)^2} + \frac{\lambda _c}{S^2}\Big]\Big[\lambda ^2 _c - \lambda _c\Big]-\Big[\frac{1}{S}(\lambda ^2 _c - \lambda _c)\Big]^2.
\end{align}
Eq. (\ref{eq:quad_S_concave}) leads to the following quadratic equation in $S$:
\beqd{
S^2(\lambda_c - 1) +2S\lambda _c + \lambda _c = 0.
}
Once more, we get to solution
\beqd{
S_i = \frac{\lambda_c +\eta _i \sqrt{\lambda _c}}{1- \lambda _c}=\frac{\sqrt{\lambda _c}(\sqrt{\lambda _c} +\eta _i)}{1- \lambda _c}.
}
We can discard $\eta _2 (-)$ in order to ensure a non-negative solution for $S _i$, leading to
\beqd
{
\label{eq:s_concave_convex}
S_{concave/convex} = \frac{\sqrt{\lambda _c}}{1-\sqrt{\lambda _c}},
}
which is exactly the dividing line as in~(\ref{eq:S_sol_sign}). Thus, we have that regions with the sign of $\psi$ and concavity/convexity regions of $(S,C)$ are identical (since $\lambda _c$ determines the same
unique $S$ in both cases and $(\lambda _c,S)$ determine a unique $C$).

A numerical calculation of this phenomenon can be easily demonstrated. We choose a square domain of $(S,C)$; calculate $(\lambda _s, \lambda _c)$ by~(\ref{eq:lambda_sc_func_SQ}) and choose the correct sign function $\Psi _i$ in order to get back $(S,C)$. Once the sign is set, we test for concavity.    

\nfigSing{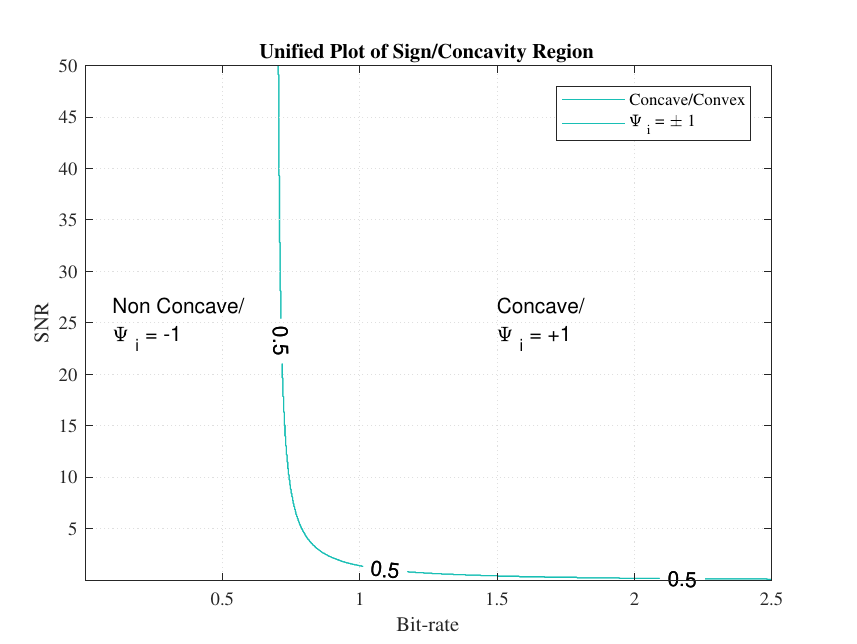}     
{Unified plot of concavity/sign regions.}
{fig:unified_plot_concave_psi}   

Fig. \ref{fig:unified_plot_concave_psi} shows that the regions of concavity and sign are identical. In this case we select the plus sign in order to get the concave solution.
To conclude, let us investigate the lower limit on $C$ using (\ref{eq:S_sol_sign}):
\beqd{
Q = \frac{1}{S}\cdot\frac{\lambda _c}{1-\lambda _c} = \frac{\sqrt{\lambda _c}}{1 +\sqrt{\lambda _c}},
}
and hence,
\beqd{
\lim _{\lambda _c \rightarrow 1} C = \lim _{\lambda _c \rightarrow 1} -log(Q) = log(2). 
}
This is the analytic derivation of the $1[bits/Hz]$ limit.

\subsection{Calculating $P_Z(Z)$ s.t. $H(Z) = C$}
\label{alg:change_induced_pmf} 
%The algorithm presented in Fig.~\ref{alg:set_pz}, is designed to calculate the pmf, $P_Z(Z)$ of the output 
This algorithm is designed to calculate the (entropy-limited deterministic) quantizer's output probability mass function, $Z$, that would meet the entropy constraint. The main idea here is to set an appropriate alphabet size $|Z|$, which is dependent upon $C$. If $e^C$ is a natural number, the alphabet size would be $|Z| = e^C$ and the probability for each output would be $e^{-C}$. If not, we define the alphabet size $|Z| = \lceil e^C \rceil$. Setting equal probability to this alphabet would yield output entropy $H(Z) > C$. At this point, we can reach the desired entropy by gradually decreasing the probability of one of the outcomes, say $Z = 1$, and increasing (uniformly) the others, thus reaching the desired entropy ($H(Z) = C$).

%% file: main.bbl
\begin{thebibliography}{10}
	\providecommand{\url}[1]{#1}
	\csname url@samestyle\endcsname
	\providecommand{\newblock}{\relax}
	\providecommand{\bibinfo}[2]{#2}
	\providecommand{\BIBentrySTDinterwordspacing}{\spaceskip=0pt\relax}
	\providecommand{\BIBentryALTinterwordstretchfactor}{4}
	\providecommand{\BIBentryALTinterwordspacing}{\spaceskip=\fontdimen2\font plus
		\BIBentryALTinterwordstretchfactor\fontdimen3\font minus
		\fontdimen4\font\relax}
	\providecommand{\BIBforeignlanguage}[2]{{%
			\expandafter\ifx\csname l@#1\endcsname\relax
			\typeout{** WARNING: IEEEtran.bst: No hyphenation pattern has been}%
			\typeout{** loaded for the language `#1'. Using the pattern for}%
			\typeout{** the default language instead.}%
			\else
			\language=\csname l@#1\endcsname
			\fi
			#2}}
	\providecommand{\BIBdecl}{\relax}
	\BIBdecl
	
	\bibitem{bib.dist_mimo_upper}
	A.~Sanderovich, S.~Shamai, and Y.~Steinberg, ``Distributed mimo
	receiver-achievable rates and upper bounds,'' \emph{IEEE Trans. Information
		Theory}, vol.~55, no.~10, pp. 4419--4438, Oct 2009.
	
	\bibitem{DcntrlizedProc}
	S.~Shamai, Y.~Steinberg, A.~Sanderovich, and G.~Kramer, ``Communication via
	decentralized processing,'' \emph{IEEE Trans. Information Theory}, vol.~54,
	no.~7, pp. 3008--3023, Jul 2008.
	
	\bibitem{bib.gib_information}
	A.~G. G.~Chechik, Y.~Weiss and N.~Tishbi, ``Information bottleneck for gaussian
	variables,'' \emph{Machine Learning Research}, vol.~6, pp. 165--188, Jan
	2005.
	
	\bibitem{bottleneck_explain}
	A.~Winkelbauer and G.~Matz, ``Rate-information-optimal gaussian channel output
	compression,'' \emph{48th Annual Conference on Information Sciences and
		Systems (CISS),}, no. 1-5, Aug 2014, doi: 10.1109/CISS.2014.6814120.
	
	\bibitem{bib.bottelneck_gaussian_vector}
	A.~Winkelbauer, S.~Farthofer, and G.~Matz, ``The rate-information trade-off for
	gaussian vector channels,'' \emph{IEEE International Symposium on Information
		Theory (ISIT)}, pp. 2849 -- 2853, Jul 2014.
	
	\bibitem{bib.berger}
	T.~Berger, \emph{Rate-Distortion Theory}.\hskip 1em plus 0.5em minus
	0.4em\relax Wiley Online Library, 2003.
	
	\bibitem{bib.Doburshin_Tsybakov}
	R.~Dobrushin and B.~Tsybakov, ``“information transmission with addi- tional
	noise,'' \emph{IRE Transactions on Information Theory}, vol.~8, no.~5, pp.
	293--304, 1962.
	
	\bibitem{bib.kipnis_et_al}
	A.~Kipnis, A.~Goldsmith, Y.~Eldar, and T.~Weissman, ``Distortion-rate function
	of sub-nyquist sampled gaussian sources,'' \emph{IEEE Transactions on
		Information Theory}, vol.~62, pp. 401--429, Jan 2016.
	
	\bibitem{DBLP:journals/corr/KipnisEG16}
	A.~Kipnis, Y.~Eldar, and A.~Goldsmith, ``Sampling stationary signals subject to
	bitrate constraints,'' \emph{arXiv:1601.06421}, 2016.
	
	\bibitem{h.gishj.n.pierce1968}
	H.~Gish and J.~Pierce, ``Asymptotically efficient quantization,'' \emph{IEEE
		Transactions on Information Theory}, vol.~14, no.~5, pp. 676--684, Sept 1968.
	
	\bibitem{p.nollr.zelinski1978}
	P.~Noll and R.~Zelinski, ``Bounds on quantizer performance in the low bit-rate
	region,'' \emph{IEEE Trans. Comm.}, vol.~26, pp. 300--304, Feb 1978.
	
	\bibitem{n.farvardinj.w.modestino1984}
	N.~Favardin and J.~W. Modestino, ``Optimum quantizer performance for a class of
	non-gaussian memoryless sources,'' \emph{IEEE Trans. Inform. Theory},
	vol.~30, no.~3, pp. 485--497, May 1984.
	
	\bibitem{a.homrim.pelegs.shamai2016}
	{Adi Homri}, {Michael Peleg}, and {Shlomo Shamai}, ``Oblivious processing in a
	fronthaul constrained gaussian channel,'' in \emph{{2016 IEEE International
			Conference on the Science of Electrical Engineering (ICSEE)}}, 2016.
	
	\bibitem{bib.shannon_basic}
	C.~Shannon, ``Communication in the presence of noise,'' \emph{Proc. IRE},
	vol.~37, no.~1, pp. 10--21, Jan 1949.
	
	\bibitem{siddhart2011}
	R.~Siddharth, M.~Medard, and L.~Zheng, ``Fiber aided wireless network
	architecture,'' \emph{IEEE JOURNAL ON SELECTED AREAS IN COMMUNICATIONS,},
	vol.~29, no.~6, pp. 1284--1294, Jun 2011.
	
	\bibitem{siddhart2006}
	S.~Ray, M.~M{\'e}dard, and L.~Zheng, ``A simo fiber aided wireless network
	architecture,'' in \emph{Information Theory, 2006 IEEE International
		Symposium on}.\hskip 1em plus 0.5em minus 0.4em\relax IEEE, 2006, pp.
	2904--2908.
	
	\bibitem{bib.cover_and_thomas}
	T.~M. Cover and J.~A. Thomas, \emph{Elements of Information Theory}, 10th~ed.,
	ser. Elements in Communications and Signal Processing.\hskip 1em plus 0.5em
	minus 0.4em\relax Wiley, Aug 1991.
	
	\bibitem{bib.gallager}
	R.~G. Gallager, \emph{Information Theory and Reliable Communications}.\hskip
	1em plus 0.5em minus 0.4em\relax New York: John Wiley, 1968.
	
	\bibitem{bib.inform_bottel_tishby}
	G.~Chechik, A.~Globerson, N.~Tishby, and Y.~Weiss, ``Information bottleneck for
	gaussian variables,'' \emph{Journal of Machine Learning Research}, no.~6, pp.
	165--188, 2005.
	
	\bibitem{bib.koch}
	T.~Koch, ``On the capacity of the dither-quantized gaussian channel,''
	\emph{IEEE International Symposium on Information Theory (ISIT)}, 2014,
	arxiv:1401.6787.
	
	\bibitem{bib.optimal_gib_curve}
	A.~Globerson and N.~Tishbi, ``On the optimality of the gaussian information
	bottleneck curve,'' \emph{Unpublished}, 2003, hUJI TR.
	
	\bibitem{bib.cap_under_output_quant}
	J.~Singh, O.~Dabeer, and U.~Madhow, ``Capacity of the discrete-time awgn
	channel under output quantization,'' \emph{IEEE International Symposium on
		Information Theory}, pp. 1218--1222, Jul 2008.
	
	\bibitem{g.kindlerr.o'donnelld.witmer2016}
	G.~Kindler, R.~O'Donnell, and D.~Witmer, ``Remarks on the most informative
	function conjecture at fixed mean,'' \emph{arXiv:1506.03167}, 2016.
	
	\bibitem{bib.hans}
	H.~S. Witsenhausen and A.~D. Wyner, ``A conditional entropy bound for a pair of
	discrete random variables,'' \emph{IEEE Trans. Information Theory}, vol.~21,
	no.~5, pp. 493--501, Sept 1975.
	
	\bibitem{DBLP:journals/corr/HomriPS15}
	{Adi Homri}, {Michael Peleg}, and {Shlomo Shamai}, ``Oblivious processing in a
	fronthaul constrained gaussian channel,'' \emph{arXiv:1510.08202}, 2015.
	
	\bibitem{bib.ib_method}
	N.~Tishbi, F.~Pereira, and W.~Bialek, ``Information bottleneck method,''
	\emph{Proc. 37th Allerton Conf. on Communication, Control and Computing}, pp.
	368--377, Sept 1999.
	
	\bibitem{bib.immse}
	D.~Guo, S.~Shamai, and S.~Verdu, ``The interplay between information ans
	estimation measures,'' \emph{Foundation and Trends in Signal Processing},
	vol.~6, no.~4, pp. 243--429, 2012.
	
	\bibitem{bib.gelfand}
	I.~M. Gelfand and S.~V. Fomin, \emph{Calculus of Variations}.\hskip 1em plus
	0.5em minus 0.4em\relax Englewood Cliffs, NJ: Prentice-Hall, 1963.
	
	\bibitem{s.hassanpourd.wuebbena.dekorsy2017}
	S.~Hassanpour, D.~Wuebben, and A.~Dekorsy, ``Overview and investigation of
	algorithms for the information bottleneck method,'' \emph{SCC}, 2017, 11th
	International ITG Conference on Systems, Communications and Coding.
	
	\bibitem{jamesmassey1976}
	J.~Massey, ``All signal sets centered about the origin are optimal at low
	energy-to-noise ratios on the awgn channel,'' \emph{ISIT}, Jun 1976.
	
	\bibitem{l.brandenburga.wyner1974}
	L.~Brandenburg and A.~Wyner, ``Capacity of the gaussian channelwith memory: The
	multivariate case,'' \emph{Bell System Technical Journal}, vol.~53, pp.
	745--779, May 1974.
	
	\bibitem{j.mietznerr.schoberl.lampew.h.gerstackerp.a.hoeher2009}
	J.~Mietzner, R.~Schober, L.~Lampe, W.~H. Gerstacker, and P.~A. Hoeher,
	``Multiple-antenna techniques for wireless communications - a comprehensive
	literature survey,'' \emph{IEEE Communications Surveys \& Tutorials},
	vol.~11, no.~2, 2009.
	
\end{thebibliography}
